\documentclass{IEEEtran}
\usepackage[T1]{fontenc}
\usepackage[latin9]{inputenc}
\usepackage{verbatim}
\usepackage{float}
\usepackage{amsmath}
\usepackage{amsthm}
\usepackage{amssymb}
\usepackage{graphicx}

\makeatletter

\floatstyle{ruled}
\newfloat{algorithm}{tbp}{loa}
\providecommand{\algorithmname}{Algorithm}
\floatname{algorithm}{\protect\algorithmname}

\theoremstyle{plain}
\newtheorem{thm}{\protect\theoremname}
\theoremstyle{definition}
\newtheorem{defn}[thm]{\protect\definitionname}
\theoremstyle{plain}
\newtheorem{lem}[thm]{\protect\lemmaname}
\theoremstyle{plain}
\newtheorem{cor}[thm]{\protect\corollaryname}
\theoremstyle{plain}
\newtheorem{prop}[thm]{\protect\propositionname}

\usepackage{cite}
\usepackage{tikz}
\usepackage{pgf}\usetikzlibrary{calc,arrows}
\usepackage{algorithm}
\usepackage{algpseudocode}

\providecommand{\lemmaname}{Lemma}
\providecommand{\theoremname}{Theorem}

\providecommand{\corollaryname}{Corollary}
\providecommand{\definitionname}{Definition}

\providecommand{\propositionname}{Proposition}

\makeatother

\providecommand{\corollaryname}{Corollary}
\providecommand{\definitionname}{Definition}
\providecommand{\lemmaname}{Lemma}
\providecommand{\propositionname}{Proposition}
\providecommand{\theoremname}{Theorem}

\begin{document}
\title{Rate-Energy Balanced Precoding Design for SWIPT based Two-Way Relay
Systems}
\author{Navneet Garg, \IEEEmembership{Member, IEEE,} Junkai Zhang, \IEEEmembership{Student Member, IEEE,}
and Tharmalingam Ratnarajah, \IEEEmembership{Senior Member, IEEE}\thanks{Authors are with Institute for Digital Communications, The University
of Edinburgh, Edinburgh, EH9 3FG, UK (e-mails: \{ngarg, jzhang15,
t.ratnarajah\}@ed.ac.uk).\protect \\
This work was supported by the UK Engineering and Physical Sciences
Research Council (EPSRC) under grant number EP/P009549/1.}}
\maketitle
\begin{abstract}
Simultaneous wireless information and power transfer (SWIPT) technique
is a popular strategy to convey both information and RF energy for
harvesting at receivers. In this regard, we consider a two-way relay
system with multiple users and a multi-antenna relay employing SWIPT
strategy, where splitting the received signal leads to a rate-energy
trade-off. In literature, the works on transceiver design have been
studied using computationally intensive and suboptimal convex relaxation
based schemes. In this paper, we study the balanced precoder design
using chordal distance (CD) decomposition, which incurs much lower
complexity, and is flexible to dynamic energy requirements. It is
analyzed that given a non-negative value of CD, the achieved harvested
energy for the proposed balanced precoder is higher than that for
the perfect interference alignment (IA) precoder. The corresponding
loss in sum rates is also analyzed via an upper bound. Simulation
results add that the IA schemes based on mean-squared error are better
suited for the SWIPT maximization than the subspace alignment-based
methods.
\end{abstract}


\section{Introduction}

With the increasing demands from a large number of devices in the
5G and beyond wireless networks, severe interference and unnecessary
power consumption are inevitable \cite{7782396}. Due to this fact,
their key performance metrics (e.g., sum rate and bit error rate)
are restricted, especially for battery operated devices \cite{9229448}.
A satisfactory solution is to harvest energy from the received RF
signal to provide a stable and long-term power supplement \cite{6230862}.
The experimental results in \cite{6513298} show that a few microwatts
of RF power can be harvested from broadcasting signals of TV stations
located several kilometers away. Thus, wireless energy harvesting
(EH) system has been employed for energy-constrained devices, such
as implantable sensors and smart wearables \cite{6957150}. Further,
since RF signals also carry information in wireless networks, simultaneous
information and power transfer (SWIPT) technology has attracted great
attention in many different scenarios \cite{Valenta2014108}.

One such scenario of interest is a two-way relay (TWR) system for
relaying information between users located at different sides of the
relay. Under the full-duplex (FD) operation at the relay node, studies
\cite{8052090,8234646,7378840,7587417,8063885} focus only on the
sum rate maximization, since self-interference cancellation requires
active circuits, causing more power consumption. Thus, the SWIPT-based
relay systems have been widely studied under the half-duplex (HD)
operation in different scenarios such as non-orthogonal multiple access
(NOMA) \cite{8636496}, mmWave with hybrid precoding \cite{8611204},
massive MIMO \cite{8048547}, wireless edge caching \cite{8474322},
Internet-of-things (IoT) \cite{8474311}, cognitive-radio networks
\cite{8000346}, secrecy systems \cite{7462485}, unmanned aerial
vehicle (UAV) \cite{8611204} and more \cite{6951347,8628978}. In
these systems, two-way relay design is presented with variations such
as single/multi-relay systems \cite{8811500}, single-hop/multi-hop
systems \cite{8352656}, and with amplify-and-forward (AF)/decode-and-forward
(DF) relaying \cite{7247664}.

Among these works, a brief review of single-hop AF relaying approaches
is given as follows. In \cite{9229448}, for the multiple-input multiple-output
(MIMO) SWIPT-based AF-TWR systems, hybridized power-time splitting
ratios and precoders are obtained via the maximization of convexified
bounds on the sum rate. In \cite{9208794,9081921}, transceivers and
splitters are designed via semi-definite relaxation (SDR) based convex
problems for finite constellation symbols. For DF relaying in \cite{7247664},
power allocation and splitting ratios are computed via the formulated
convex problem. In \cite{7981373}, both source and relay are designed
using successive convex approximation. In \cite{7880771,8325514,7876801},
energy efficiency is optimized with respect to joint source and relay
power allocation via relaxed and convexified objective. In \cite{8649668},
asymptotic bit error rate is analyzed for space-shift-keying modulation,
while \cite{7740048} analyzes outage probability to verify the similar
diversity order for SWIPT as in the non-SWIPT case. In \cite{8337780,8361446},
dynamic and asymmetric splitting ratios are computed via iterative
Dinkelbach-based algorithm to solve non-convex problem. In these works,
first, transceivers are designed in a suboptimal manner. Second, they
focus only on sum rate, although SWIPT model is adopted for energy
harvesting. Thus, in the SWIPT model, an effective rate-energy trade-off
needs more investigation.

Further, since two-way relay causes interference at receivers, interference
management approaches are also studied. In \cite{6880863} for DF
relaying, non-cooperative game is formulated to utilize the relay
resource, where each user maximizes its own rate in an interference
channel. Interference among the SWIPT-relay assisted channels is mitigated
via NOMA in \cite{9037234} and a closed form transmitter design is
provided for a fixed split-ratio. In \cite{7920315}, beamforming
is obtained for different relaying protocols (AF/DF) via suboptimal
convex relaxations. In \cite{7559785}, joint source and relay transceivers
are designed to maximize the energy efficiency via convex approaches.
Thus, for these works also, the suboptimal transceiver design and
the focus on sum rates somewhat defeats the purpose of SWIPT operation.
Regarding interference management, interference alignment (IA) has
been a popular method for a decade \cite{6567868,6784515,7850961,8280560}.
In the typical IA for MIMO interference channels (ICs), precoders
for two interfering transmitters are designed to align the interfering
signals at the receiver subspace, and the receiver can use a decoder
(a linear combiner) to null the aligned interference \cite{Cadambe2008,9097459}.
The following works uses IA in the SWIPT-relay system to mitigate
the interference. In \cite{9042895}, IA is used to improve the harvested
energy, while keeping transmissions secure by introducing artificial
noise. In \cite{8226830}, a two-stage splitting scheme is derived
with IA to maximize the sum rate. For MIMO broadcast channels \cite{7572098},
block diagonalization-based method to get the improved sum rates.
For massive MIMO system, asymptotic SINR is analyzed for a signal-space-alignment
method in \cite{8356679}. Therefore, the study of IA with a general
two-way relay system is lacking towards a low-complexity design with
better rate-energy trade-offs.

\subsection{Contributions}

In this paper, we consider the system with a SWIPT based two-way relay
serving multiple user nodes, who wish to communicate to other users
via the relay under the HD operation. First, to avoid processing at
the relay to reduce power consumption, the AF relaying is adopted,
using which precoders and decoders are computed by modifying an IA
algorithm based on minimum mean-squared error (MMSE) \cite{9097459}.
With the perfect IA precoders obtained, we provide a systematic process
to get the balanced precoder using chordal distance (CD) decomposition
to improve the harvested energy. Via rate-energy trade-off, it is
observed that improvement in harvested energy leads to reduction in
sum rates, which can be decided using the CD value. Maximum harvested
energy-based precoders are also obtained and the corresponding rate
loss is analyzed. In simulations, rate-energy regions are plotted
for different precoding methods for different CD values. These results
show that a better rate-energy trade-off can be obtained, as compared
to the other transceiver designs. The contributions can be summarized
as follows: 
\begin{itemize}
\item TWR-IA algorithm: Since the effective end-to-end channel includes
the relay processing matrix, it is challenging to find an optimum
precoding scheme, as in an iterative IA algorithm the effective channel
varies with each iteration. From our experiments, it turns out that
AF provides the best end-to-end sum rate. Further, since an IA method
cannot be directly applied here, the required modifications lead to
the different precoders and decoders expressions, which in turn are
formulated into the TWRIA algorithm. 
\item Balanced precoding: In order to improve the harvested energy at the
relay, the CD decomposition is used to compute the balanced precoder
for rate-energy trade-off, which can provide higher energy while keeping
the expected rate-loss constant proportional to the CD value. In other
words, the desired sum rate reduction can be specified via the CD
values and the splitting ratio to obtain higher energies. 
\item Analysis and simulations: Maximum achievable energy, rate loss, and
harvested energy bounds are obtained to verify and analyze the proposed
balanced precoding. Further, simulations with two different IA methods
show the better sum rates for the TWRIA algorithm, and the better
trade-offs for the balanced precoding with respect to different CD
values. 
\end{itemize}
The rest of the paper is organized as follows: the symmetric two-way
relay IC system model is given in Section \ref{ii}, followed by an
energy optimized precoding method and a rate-energy balanced precoding
design algorithm in Section \ref{iii} and \ref{iv}, respectively.
Simulation results are shown in Section \ref{v}. A brief conclusion
of this work is presented in Section \ref{vi}.

\subsubsection*{Notations}

$\mathcal{B}$, $\mathbf{B},\mathbf{b}$, $b$ represent a set, a
matrix, a vector, and a scalar, respectively. The notations $\mathbf{B}^{H}$,
$\mathbf{B}^{-1}$, $\mathbf{B}(m,n)$, $\lVert\mathbf{B}\rVert_{F}$,
$\|\mathbf{B}\|$, $|\mathbf{B}|$, $\Re\text{tr}(\mathbf{B})$ and
$\nu_{1:b}[\mathbf{B}]$ are the Hermitian transpose, the inverse
of $\mathbf{B}$, the $(m,n)^{th}$ value of the matrix $\mathbf{B}$
(also denoted by $\left[\mathbf{B}\right]_{m,n}$), the Frobenius
norm, spectral norm, the determinant, the real part of trace, and
$\nu_{1:b}[\mathbf{B}]$ denotes the first $b$ dominant eigenvectors
of $\mathbf{B}$, respectively. $\lVert\mathbf{b}\rVert_{2}$ denotes
the $l_{2}$-norm of $\mathbf{b}$. $\mathrm{Cov}(\mathbf{b})=\mathbb{E}\left\{ \mathbf{b}\mathbf{b}^{H}\right\} $
is the covariance matrix of zero mean vector $\mathbf{b}$, where
$\mathbb{E}\{\cdot\}$ is the expectation operator. $\mathcal{D}(\mathbf{B}_{1},\mathbf{B}_{2})$
denotes a block diagonal matrix with $\mathbf{B}_{1}$ and $\mathbf{B}_{2}$
as its block diagonal components. $\mathcal{CN}(b,\mathbf{B})$ represents
a circularly symmetric complex Gaussian random vector with mean $b$
and covariance matrix $\mathbf{B}$. $\mathbb{O}[\cdot]$ denotes
the orthonormal operator, can be obtained from QR-decomposition, and
$\mathbf{I}_{K}$ is a $K\times K$ identity matrix.

\section{System Model\label{ii}}

\begin{figure}[!t]
\centering \includegraphics[width=1\columnwidth]{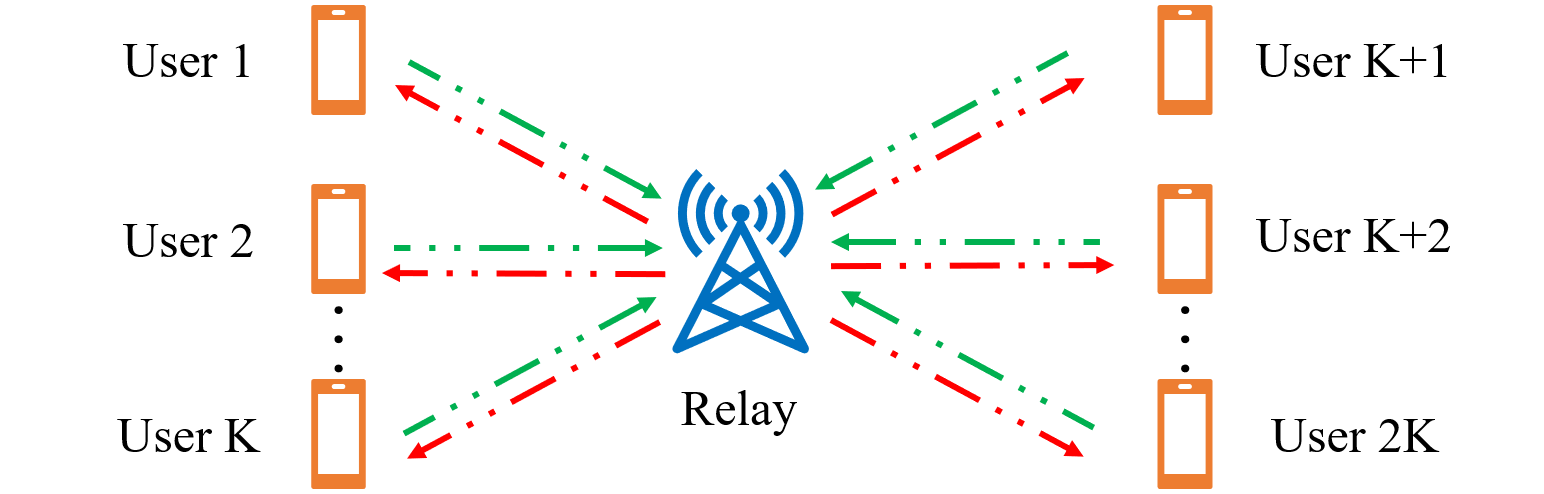} \caption{Illustration of $K$-user pairs in the TWR system.\label{model}}
\end{figure}

Consider a symmetric TWR IC \cite{5956515} with $2K$ user nodes,
as shown in Figure \ref{model}. 
Each of the $2K$ nodes equipped with $M$ antennas wants to transmit
$d$ data streams to its paired user node via a $R$-antenna TWR.
The destination (or source) of the $k^{th}$ source (or destination)
is indexed by $k'=\text{mod}\left(k+K-1,K\right)+1$, for $k=1,\ldots,2K$.
We assume direct links between sources and destinations are unavailable,
which usually occurs when the direct link is blocked due to long-distance
path loss or obstacles \cite{8325514,8675436}. We use $(M,R,d)^{2K}$
to denote the setting of this symmetric TWR IC. With the HD relaying,
the communication period is divided into two phases, namely, multiple
access (MAC) and broadcast (BC) phases. For simplicity, the energy-constrained
relay node is operated under the AF mode \cite{9233408}. As in \cite{8325514},
we assume that the channel varies slowly enough so that it can be
perfectly estimated by training sequences or feedback.

\subsection{MAC phase}

In the MAC phase, each of $2K$ users transmits its $d$ data streams
to the relay, leading to the received signal equation at the relay
as 
\begin{equation}
\mathbf{y}_{r}=\sum_{j=1}^{2K}\mathbf{H}_{rj}\mathbf{V}_{j}\mathbf{s}_{j}+\mathbf{z}_{r},
\end{equation}
where $\mathbf{H}_{rj}\in\mathbb{C}^{R\times M}$ denotes the wireless
MIMO channel matrix from the $j^{th}$ user to the relay node; the
matrix $\mathbf{V}_{j}\in\mathcal{G}_{M,d}$ is the orthonormal precoder
at the $j^{th}$ user such that $\mathbf{V}_{j}^{H}\mathbf{V}_{j}=\mathbf{I}_{d},\forall j$;
the vector $\mathbf{s}_{k}\in\mathbb{C}^{d\times1}$ represents the
uncorrelated transmit data i.e., $\mathbb{E}\left\{ \mathbf{s}_{k}\mathbf{s}_{k}^{H}\right\} =\frac{P_{k}}{d}\mathbf{I}_{d}$
with $P_{k}$ being the total transmit power at the $k^{th}$ user
node; and $\mathbf{z}_{r}\sim\mathcal{CN}\left(0,\sigma_{R}^{2}\mathbf{I}_{R}\right)$
is the Gaussian noise. Entries of the channel matrix $\mathbf{H}_{rj}$
are assumed to have zero mean and variance $\mathbb{E}\left|\mathbf{H}_{rj}(m,n)\right|^{2}=\beta_{rj},\forall m,n$.
Thus, the relay forwards $2Kd$ data streams. Conventionally, the
condition $R\geq2Kd$ is required. However, with IA, $Kd$ antennas
at the relay are sufficient \cite{5956515}.

\subsection{Harvesting Energy}

After the MAC phase, the relay splits the received signal into two
flows: one part goes for EH, while the remaining is will be forwarded
in the BC phase for information decoding (ID) at users. The received
signal $\mathbf{y}_{r}$ is fed into a power splitter with a PS ratio
$\rho\in\left[0,1\right]$, denoting the portion for harvesting energy,
\begin{align}
\mathbf{y}_{r}^{EH} & \approx\sqrt{\rho}\mathbf{y}_{r}=\sqrt{\rho}\left(\sum_{j=1}^{2K}\mathbf{H}_{rj}\mathbf{V}_{j}\mathbf{s}_{j}+\mathbf{z}_{r}\right),
\end{align}
where in the above approximation, the noise introduced by the splitter
is negligible compared to the received signal strength, and hence
ignored. The corresponding average harvested energy at the relay can
be expressed as 
\begin{subequations}
\begin{align}
Q_{r} & =\zeta\mathbb{E}\left\{ \left\Vert \mathbf{y}_{r}^{EH}\right\Vert _{2}^{2}\right\} ,\\
 & \approx\zeta\rho\sum_{j=1}^{2K}\frac{P_{j}}{d}\left\Vert \mathbf{H}_{rj}\mathbf{V}_{j}\right\Vert _{F}^{2},\label{eq:max_EH-Qk}
\end{align}
\end{subequations}
 where $\zeta\in\left[0,1\right]$ represents the RF-to-electrical
conversion efficiency. Note that the noise power $\zeta\rho\sigma_{r}^{2}R$
is negligible and constant, and hence is omitted in the above equation.

\subsection{BC Phase}

The received signal at the power splitter for ID experiences an additional
circuit noise due to non-ideal splitters, non-ideal RF-baseband conversion
and thermal noise \cite{7230306}. Thus, the signal for ID at the
relay can be expressed as 
\begin{subequations}
\begin{align}
\mathbf{y}_{r}^{ID} & =\sqrt{\bar{\rho}}\mathbf{y}_{r}+\mathbf{w}_{r}\\
 & =\sqrt{\bar{\rho}}\left(\sum_{j=1}^{2K}\mathbf{H}_{rj}\mathbf{V}_{j}\mathbf{s}_{j}+\mathbf{z}_{r}\right)+\mathbf{w}_{r},
\end{align}
\end{subequations}
 where $\bar{\rho}=1-\rho$ and $\mathbf{w}_{r}\sim\mathcal{CN}\left(\mathbf{0},\delta^{2}\mathbf{I}_{R}\right)$.
It can be noted that the above equation results in an effective noise
$\tilde{\mathbf{w}}_{r}=\mathbf{z}_{r}+\frac{\mathbf{w}_{r}}{\sqrt{\bar{\rho}}}\sim\mathcal{CN}\left(\mathbf{0},\sigma_{ID}^{2}\mathbf{I}_{R}\right)$,
where $\sigma_{ID}^{2}=\sigma_{R}^{2}\left(1+\frac{\delta^{2}}{\bar{\rho}\sigma_{R}^{2}}\right)$.

Next, in the BC phase, the relay first precodes the signal using the
matrix $\mathbf{G}\in\mathbb{C}^{R\times R}$ satisfying the transmit
power constraint at the relay. Then, the relay broadcasts the noisy-precoded
signal to user destinations. At the $k^{th}$ user, the received signal
is written as 
\begin{subequations}
\begin{align}
\mathbf{y}_{k} & =\mathbf{H}_{kr}\mathbf{G}\mathbf{y}_{r}^{ID}+\mathbf{n}_{k}\\
 & =\sqrt{\bar{\rho}}\mathbf{H}_{kr}\mathbf{G}\sum_{j=1}^{2K}\mathbf{H}_{rj}\mathbf{V}_{j}\mathbf{s}_{j}+\sqrt{\bar{\rho}}\mathbf{H}_{kr}\mathbf{G}\tilde{\mathbf{w}}_{r}+\mathbf{n}_{k}\\
 & =\sqrt{\bar{\rho}}\mathbf{H}_{kr}\mathbf{G}\mathbf{H}_{rk'}\mathbf{V}_{k'}\mathbf{s}_{k'}+\sqrt{\bar{\rho}}\mathbf{H}_{kr}\mathbf{G}\mathbf{H}_{rk}\mathbf{V}_{k}\mathbf{s}_{k}\\
 & \quad+\sqrt{\bar{\rho}}\mathbf{H}_{kr}\mathbf{G}\sum_{j\neq k,k'}\mathbf{H}_{rj}\mathbf{V}_{j}\mathbf{s}_{j}+\sqrt{\bar{\rho}}\mathbf{H}_{kr}\mathbf{G}\tilde{\mathbf{w}}_{r}+\mathbf{n}_{k},
\end{align}
\end{subequations}
 where $\mathbf{H}_{kr}\in\mathbb{C}^{M\times R}$ denotes the wireless
MIMO channel matrix from the relay node to the $k^{th}$ user with
zero mean and variance $\mathbb{E}\left|\mathbf{H}_{kr}(m,n)\right|^{2}=\beta_{kr},\forall m,n$;
the vector $\mathbf{n}_{k}\sim\mathcal{CN}\left(\mathbf{0},\sigma^{2}\mathbf{I}_{M}\right)$
is the Gaussian noise at the receiver. The above equation respectively
consists of the desired signal term, the self-interference component,
the co-channel interference, the relayed noise, and the received noise.
For the information retrieval and to mitigate the interference, the
$k^{th}$ node deducts the self-interference term and then employs
a combiner matrix $\mathbf{U}_{k}$ as 
\begin{equation}
\hat{\mathbf{y}}_{k}=\mathbf{U}_{k}^{H}\left(\mathbf{y}_{k}-\sqrt{\bar{\rho}}\mathbf{H}_{kr}\mathbf{G}\mathbf{H}_{rk}\mathbf{V}_{k}\mathbf{s}_{k}\right),\label{yk}
\end{equation}
where for simplicity, $\mathbf{U}_{k}$ is assumed to be an orthonormal
matrix, i.e., $\mathbf{U}_{k}^{H}\mathbf{U}_{k}=\mathbf{I}_{d}$.

\subsubsection{IA feasibility}

Let $\mathbf{H}_{kj}=\mathbf{H}_{kr}\mathbf{G}\mathbf{H}_{rj}$. For
IA, the set of precoders and combiners need the satisfy the following
equations 
\begin{subequations}
\begin{align}
\mathbf{U}_{k}^{H}\mathbf{H}_{kj}\mathbf{V}_{j}=\mathbf{0}, & \forall j\neq k,k',\label{eq:ia1}\\
\text{rank}\left(\mathbf{U}_{k}^{H}\left[\mathbf{H}_{kk}\mathbf{V}_{k},\mathbf{H}_{kk'}\mathbf{V}_{k'}\right]\right) & \geq d,\forall k.\label{eq:ia2}
\end{align}
\end{subequations}

The first equation ensures the interference terms are zero at all
receivers, while the second terms ensures the availability of at least
$d$-dimensions for the decoding of the desired signal. The desired
signal can occupy the same subspace as the self-interference signal,
since the known self-signal term can be removed by subtraction as
in \eqref{yk}. Note that each relay receives $2Kd$ data streams,
thus without IA, the relay requires at least $2Kd$ antennas ($R\geq2Kd$).
However, for less number of antennas at relay, i.e., if $Kd\leq R<2Kd$,
each node's precoder should be aligned with the desired signal subspace,
that is, 
\begin{equation}
\text{span}\left(\mathbf{H}_{jk}\mathbf{V}_{k}\right)=\text{span}\left(\mathbf{H}_{jk'}\mathbf{V}_{k'}\right),\forall j,k.
\end{equation}
It can be seen from the above equation \eqref{eq:ia1}-\eqref{eq:ia2}
that for a fixed relay matrix $\mathbf{G}$, the above model is analogous
to an interference channel $\left(M\times M,d\right)^{2K}$ with the
equivalent channel matrices $\mathbf{H}_{kj}$, $\forall k,j$. For
this analogous IC, one needs to satisfy the necessary proper system
condition in \cite{5466114}, or the guaranteed IA-feasibility condition
in \cite{9097459}. From the above equations (for all $k,j$), by
setting the number of equations $\left(d^{2}2K(2K-2)\right)$ less
than or equal to the number of variables $\left(4K(M-d)d\right)$,
we arrive at the necessary condition 
\begin{equation}
M\geq Kd.
\end{equation}

\subsubsection{Sum rate}

At the $k^{th}$ destination node, the resulting rate can be written
as $R_{k}=\frac{1}{2}\times$
\[
\log_{2}\Bigg|\mathbf{I}_{d}+\frac{\bar{\rho}P_{k'}}{d}\bar{\mathbf{H}}_{kk'}\bar{\mathbf{H}}_{kk'}^{H}\left(\mathbf{N}_{k}+\sum_{j\neq k,k'}\frac{\bar{\rho}P_{j}}{d}\bar{\mathbf{H}}_{kj}\bar{\mathbf{H}}_{kj}^{H}\right)^{-1}\Bigg|,
\]
where $\bar{\mathbf{H}}_{kj}=\mathbf{U}_{k}^{H}\mathbf{H}_{kj}\mathbf{V}_{j}$;
$\mathbf{N}_{k}=\mathbf{U}_{k}^{H}\mathbf{C}_{k}\mathbf{U}_{k}$ with
$\mathbf{C}_{k}$ being the effective noise covariance matrix given
as 
\begin{subequations}
\begin{align*}
\mathbf{C}_{k} & =\text{Cov}\left(\sqrt{\bar{\rho}}\mathbf{H}_{kr}\mathbf{G}\tilde{\mathbf{w}}_{r}+\mathbf{n}_{k}\right)\\
 & =\bar{\rho}\sigma_{ID}^{2}\mathbf{H}_{kr}\mathbf{G}\mathbf{G}^{H}\mathbf{H}_{kr}^{H}+\sigma^{2}\mathbf{I}_{M}.
\end{align*}
\end{subequations}

If the interference is perfectly canceled, i.e., $\bar{\mathbf{H}}_{kj}=\mathbf{0}$,
the rate is constrained by the noise component forwarded by the relay
as 
\[
R_{k,per}=\frac{1}{2}\log_{2}\left|\mathbf{I}_{d}+\bar{\rho}\frac{P_{k'}}{d}\bar{\mathbf{H}}_{kk'}\bar{\mathbf{H}}_{kk'}^{H}\mathbf{N}_{k}^{-1}\right|.
\]
To obtain the limits on harvested energy, we first present both the
rate and the energy optimized precoding with its analysis in the following.

\section{Precoding for Rate or energy limits\label{iii}}

In this section, we first provide a brief overview of the modified
IA algorithm for the TWR system. Since the focus of the paper is SWIPT
schemes, the details of the TWRIA algorithm are delegated to the Appendix\ref{APPIA}.
After the rate-optimized precoding, the precoders achieving the maximum
harvested energy are derived and the expected rate-loss upper bound
is analyzed. Subsequently, the definition of CD and its properties
are explained to introduce the balance precoding.

\subsection{Rate-optimized precoding: TWRIA algorithm}

Conventional IA methods for interference mitigation are suited to
interference channels. In TWR system, due to the presence of relay,
the channel matrix depends on the choice of relay processing matrix
$\mathbf{G}$ with relay transmit power constraint. Owing to this
dependence, the effective channel matrix $\mathbf{H}_{kj}$ varies
in each iteration for a conventional iterative IA algorithm, leading
to a much higher computationally overhead and slower convergence speed.
Therefore, the TWR variant of the MSE-based IA method \cite{9097459}
is derived in the Appendix\ref{APPIA}. Similar to the IA method in
\cite{9097459}, the TWRIA is an iterative algorithm, where precoders
and combiners are alternately updated using the corresponding set
of expressions derived in \eqref{eq:Vk} and \eqref{eq:Uk}, respectively.

In TWR-IA algorithm, the relay matrix $\mathbf{G}$ is chosen proportional
to an identity matrix, $\mathbf{G}=\alpha\mathbf{I}_{R}$, due to
following reasons.
\begin{itemize}
\item IA conditions: The first interference alignment condition $\mathbf{U}_{k}^{H}\mathbf{H}_{kr}\mathbf{G}\mathbf{H}_{rj}\mathbf{V}_{j}=\mathbf{0},\forall j\neq k,k',$
is unaffected by the choice of $\mathbf{G}$. At the end of IA algorithm,
this product is close to zero via the matrices $\mathbf{U}_{k}$ and
$\mathbf{V}_{k}$, irrespective of the value of $\mathbf{G}$. Moreover,
due to the presence of $\mathbf{U}_{k}$ and $\mathbf{V}_{k}$ at
both sides of the effective channel $\frac{\mathbf{H}_{kr}\mathbf{G}\mathbf{H}_{rj}}{\|\mathbf{G}\|_{F}}$
for all $k,j$, depending on $\frac{\mathbf{G}}{\|\mathbf{G}\|_{F}}$,
the variables $\mathbf{U}_{k}$ and $\mathbf{V}_{k}$ will be optimized
accordingly via the TWRIA algorithm. In other words, $\mathbf{U}_{k}$
and $\mathbf{V}_{k}$ are the main matrices deciding the structure
of the product $\mathbf{U}_{k}^{H}\mathbf{H}_{kr}\frac{\mathbf{G}}{\|\mathbf{G}\|_{F}}\mathbf{H}_{rj}\mathbf{V}_{j}$
for any $k,j$, and the final MSE-values, rather than the structure
of $\mathbf{G}$. The only part of $\mathbf{G}$, that affects the
MSE or sum rate performance is its weight factor $\alpha$, which
is chosen according to the relay transmit power constraint.
\item Diagonalization: Since the matrix $\mathbf{G}$ acts as a trio of
a receiver, an amplifier and a transmitter, we can write via SVD as
$\mathbf{G}=\mathbf{G}_{T}\Lambda_{G}\mathbf{G}_{R}$, where $\mathbf{G}_{T}$
and $\mathbf{G}_{R}$ are $R\times R$ orthonormal matrices acting
as relay transmit precoder and receive decoder; and $\Lambda_{G}$
is a $R\times R$ diagonal matrix with non-negative entries for relay
amplification. Note that an square orthonormal matrix is unitary matrix,
i.e., $\mathbf{G}_{T}^{H}\mathbf{G}_{T}=\mathbf{G}_{T}\mathbf{G}_{T}^{H}=\mathbf{G}_{R}^{H}\mathbf{G}_{R}=\mathbf{G}_{R}\mathbf{G}_{R}^{H}=\mathbf{I}$,
i.e., it does not contribute to the interference alignment or the
transmit power changes. Therefore, without loosing optimality, it
is sufficient to consider $\mathbf{G}=\Lambda_{G}$. Further, for
the received signal at relay i.e., $\sum_{j=1}^{2K}\mathbf{H}_{rj}\mathbf{V}_{j}\mathbf{s}_{j}$,
each entry of this vector contains all channels and data streams.
Note that at the relay, each of $2Kd$ data streams are equally important
to forward. Thus, unequal power allocation $\left(\Lambda_{G}\right)$
to minimize the system objective, such as MSE or sum rate, will provide
diagonal values in $\Lambda_{G}$ proportional to the small-scale
fading variations, summed across all users. As the number of users
or antennas increases, these small-scale variations reduce. Therefore,
for simplicity and without loosing much optimality,  the relay matrix
$\Lambda_{G}$ is relaxed to $\alpha\mathbf{I}_{R}$. 
\end{itemize}

Regarding the convergence of TWRIA algorithm, it can be seen that
the total MSE is jointly convex with respect to precoders and combiners
(see \cite{9097459}). Thus, the algorithm converge globally and is
demonstrated via simulation results. The computational complexity is
same as an conventional IA algorithm, where the number of iterations
for convergence depends on SNR. Further, it can be noted that the
TWRIA algorithm can be designed independent of splitting operation
for SWIPT. This important feature along with the low-complexity balanced
precoding allows the use of dynamic splitting ratios in order to dynamically
satisfy rate and energy constraints. To provide the SWIPT functionality,
CD decomposition is discussed in the next section. To analyze the
rate-energy trade-off, it is important to quantify the maximum harvested
energy achievable, which is given as follows.

\subsection{Energy-optimized precoding \label{subsec:Maximum-EH-Precoding}}

The problem of maximizing the harvested energy at the relay with respect
to precoders, subject to orthogonality constraint on the precoders,
can be written as 
\begin{subequations}
\begin{align}
\left\{ \mathbf{V}_{j}^{EH},\forall j\right\} =\arg\max_{\mathbf{V}_{j},\forall j} & \zeta\rho\sum_{j}\frac{P_{j}}{d}\|\mathbf{H}_{rj}\mathbf{V}_{j}\|_{F}^{2}\\
\text{subject to } & \|\mathbf{V}_{j}\|_{F}^{2}\leq d,\forall j.
\end{align}
\end{subequations}
 The above problem can be decoupled, and the solution of the $j^{th}$
precoder $\mathbf{V}_{j}$ can be obtained by the dominant eigenvectors
of $\mathbf{H}_{rj}^{H}\mathbf{H}_{rj}$ corresponding to $d$ maximum
eigenvalues, i.e., 
\begin{subequations}
\begin{align}
\mathbf{V}_{j}^{EH} & =\arg\max_{\|\mathbf{V}_{j}\|_{F}^{2}\leq d}\text{tr}\left(\mathbf{V}_{j}^{H}\mathbf{H}_{rj}^{H}\mathbf{H}_{rj}\mathbf{V}_{j}\right)\label{eq:max_EH-Vj}\\
 & =\nu_{1:d}\left[\mathbf{H}_{rj}^{H}\mathbf{H}_{rj}\right]=\mathbf{W}_{j}^{[1]},
\end{align}
\end{subequations}
 where $\mathbf{W}_{j}^{[1]}$ is computed via the eigenvalue decomposition
(EVD), i.e., 
\begin{equation}
\mathbf{H}_{j}^{H}\mathbf{H}_{j}=\sum_{k}\bar{\rho}_{k}\mathbf{H}_{kj}^{H}\mathbf{H}_{kj}=\mathbf{W}_{j}\mathbf{\Lambda}_{j}\mathbf{W}_{j}^{H},\label{eq:EVD_EH}
\end{equation}
with $\mathbf{W}_{j}=\left[\mathbf{W}_{j}^{[1]},\mathbf{W}_{j}^{[2]}\right]$
and $\mathbf{\Lambda}_{j}=\mathcal{D}\left(\lambda_{ji},i=1,\ldots,M\right)$,
such that $\lambda_{j1}\geq\cdots\geq\lambda_{jM}$ being in the descending
order. Note that $\mathbf{W}_{j}^{[1]}$ and $\mathbf{W}_{j}^{[2]}$
are orthonormal matrices of size $M\times d$ and $M\times M-d$,
respectively. To analyze the effect of the precoding scheme on sum
rates, we utilize CD and its decomposition, which are defined in the
following.

\subsection{Chordal Distance}
\begin{defn}
Let $\mathbf{V},\hat{\mathbf{V}}\in\mathbb{C}^{M\times d}$ be two
orthonormal matrices such that $\hat{\mathbf{V}}^{H}\hat{\mathbf{V}}=\mathbf{V}^{H}\mathbf{V}=\mathbf{I}_{d}$.
The CD between these matrices can be defined as 
\begin{equation}
d_{c}^{2}(\mathbf{V},\,\hat{\mathbf{V}})=\frac{1}{2}\|\mathbf{V}\mathbf{V}^{H}-\hat{\mathbf{V}}\hat{\mathbf{V}}^{H}\|_{F}^{2}=d-\|\mathbf{V}^{H}\hat{\mathbf{V}}\|_{F}^{2}.\label{eq:chord_dist_mat-1}
\end{equation}
\end{defn}
Note that the matrices $\mathbf{V}$ and $\hat{\mathbf{V}}$ represent
$d$ dimensional subspaces of $M$ dimensional space, i.e., $\mathbf{V}$
and $\hat{\mathbf{V}}$ lie on a Grassmannian manifold $\mathcal{G}_{M,d}$,
which is a collection of all such $d$ dimensional subspaces. The
CD represents the distance between the subspaces spanned by these
matrices. Thus, two orthonormal matrices that represent the same column
space will have zero CD value. The CD value between two unit-norm
vectors (say $\mathbf{v}_{1},\mathbf{v}_{2}\in\mathcal{G}_{M,1}$)
is equivalent to computing the inner-product between them, i.e., $1-\left|\mathbf{v}_{1}^{H}\mathbf{v}_{2}\right|^{2}$.
Further, given two matrices in $\mathcal{G}_{M,d}$, one matrix can
be expressed into the other one using the CD decomposition lemma from
\cite[Lemma 1]{4641957}. The following lemma states the modified
CD decomposition, where the modification comes from splitting the
null space of dimension $M-d$ into a product of two matrices. 
\begin{lem}
\label{lem:MAT-CD-decomp-1}The two matrices $\hat{\mathbf{V}}$ and
$\mathbf{V}$ (such that $\hat{\mathbf{V}}^{H}\hat{\mathbf{V}}=\mathbf{V}^{H}\mathbf{V}=\mathbf{I}_{d}$)
admits the following decomposition \cite[Lem 1]{4641957} 
\begin{equation}
\mathbf{V}=\hat{\mathbf{V}}\mathbf{X}\mathbf{Y}+\hat{\mathbf{V}}^{\text{null}}\mathbf{S}\mathbf{Z},\label{eq:cd_decomp-1}
\end{equation}
where $\mathbf{V},\,\hat{\mathbf{V}}\in\mathbb{C}^{M\times d}$, $\hat{\mathbf{V}}_{j}^{\text{null}}=\text{null}(\hat{\mathbf{V}}_{j})\in\mathbb{C}^{M-d\times d},$
$\mathbf{X}\in\mathbb{C}^{d\times d}$ and $\mathbf{S}\in\mathbb{C}^{M-d\times d}$
are orthonormal matrices, $\mathbf{Y},\,\mathbf{Z}\in\mathbb{C}^{d\times d}$
are upper triangular matrices with positive diagonal elements satisfying
\begin{subequations}
\begin{eqnarray}
\mathrm{tr}(\mathbf{Z}^{H}\mathbf{Z}) & = & d_{c}^{2}(\mathbf{V},\hat{\mathbf{V}}),\\
\mathbf{Y}^{H}\mathbf{Y} & = & \mathbf{I}_{d}-\mathbf{Z}^{H}\mathbf{Z},\label{eq:decomp-YY-1}
\end{eqnarray}
\end{subequations}
 Moreover, $\mathbf{X}$ and $\mathbf{Y}$ are distributed independently
of each other, as is the pair $\mathbf{S}$ and $\mathbf{Z}$. 
\end{lem}
\begin{IEEEproof}
A short proof is included in Appendix\ref{sec:Proof-of-CD-1} from
\cite{4641957}, including the proofs of the following corollaries. 
\end{IEEEproof}
Note that this decomposition requires $M\geq2d$, which is the case
in IA, i.e., at least $d$ dimensions for the desired signal and the
remaining for the interference. 
\begin{cor}
\label{cor:CDD1}If two sets of precoders have zero chordal distances,
the resulting rate and the harvested energy are the same. 
\end{cor}

Note that the two different orthogonal matrices with zero chordal
distance will be termed as equivalent matrices; however, they cannot
be considered as the same matrix. 
\begin{cor}
\label{cor:Given-CD}Given the CD value $z$ and an orthogonal matrix
$\mathbf{V}$. Then, in obtaining the displacement precoder (with
respect to $\mathbf{V}$) via the CD decomposition, the matrices $\mathbf{Y}$
and $\mathbf{Z}$ can be relaxed to diagonal matrices as 
\begin{equation}
\mathbf{V}_{D}=\mathbf{V}\mathbf{X}\Sigma_{Y}+\mathbf{V}^{\text{null}}\mathbf{S}\Sigma_{Z},
\end{equation}
where $\Sigma_{Y}$ and $\Sigma_{Z}$ are diagonal matrices such that
$\Sigma_{Y}^{2}=\mathbf{I}_{d}-\Sigma_{Z}^{2}$. 
\end{cor}

\subsection{Rate loss upper bound}

With the maximum EH based precoding in \eqref{eq:max_EH-Vj}, the
resultant maximum harvested energy can be written as the sum of the
first $d$ dominant eigenvalues of $\mathbf{H}_{rj}^{H}\mathbf{H}_{rj},\forall j$.
Note that the precoding in \eqref{eq:max_EH-Vj} is an independent
precoding scheme, which does not mitigate the effect of interference
terms for ID. However, the obtained precoders may partially align
the interference. This partial alignment can be measured using the
CD between the ideal IA precoders and the EH precoders, say
\begin{equation}
z_{j}^{EH}=d_{c}^{2}(\mathbf{V}_{j},\mathbf{V}_{j}^{EH}),\forall j,
\end{equation}
where $\mathbf{V}_{j},\forall j$ stands for IA-precoders. It can
be noted that the above CD represents the displacement of $\mathbf{V}_{j}^{EH}$
with respect to $\mathbf{V}_{j}$, and it does not depend on SNR values.
The more the distance, the more will be interference. Therefore, it
is essential to specify the allowable interference in the system,
which can be characterized in the following result \cite{8379356}. 
\begin{lem}
\label{lem:Rate-loss-basic}(Rate Loss Upper Bound) In the TWR system
$(M,R,d)^{2K}$, the usage of imperfect precoder instead of TWRIA
precoder at the sources incurs the rate loss $\Delta R_{k}$, whose
expected value can be upper bounded for the $k^{th}$ receiver as
$\mathbb{E}\left\{ \Delta R_{k}\right\} <$
\begin{align}
 & \frac{d}{2}\log_{2}\left[1+\frac{M_{d}\bar{\rho}\sum_{j\neq k,k'}P_{j}z_{j}\beta_{rj}}{\bar{\rho}\left(\sum_{j=1}^{2K}P_{j}\frac{\beta_{rj}\sigma^{2}}{\beta_{kr}P_{r}}+\bar{\sigma}_{kr}^{2}\right)+\bar{\delta}_{kr}^{2}}\right],\label{eq:lemmarateloss}
\end{align}
with $M_{d}=\frac{M}{d(M-d)}$, $\bar{\sigma}_{kr}^{2}=\sigma_{R}^{2}\left(1+\frac{\sigma^{2}}{\beta_{kr}P_{r}}\right)$,
$\bar{\delta}_{kr}^{2}=\delta^{2}\left(1+\frac{\sigma^{2}}{\beta_{kr}P_{r}}\right)$,
$z_{k}=\mathbb{E}d_{c}^{2}(\mathbf{V}_{k},\hat{\mathbf{V}}_{k})$
being the average CD between the IA precoder and the imperfect one. 
\end{lem}
\begin{IEEEproof}
Proof is given in Appendix\ref{subsec:Proof-of-LemmaRLUB}. 
\end{IEEEproof}
The above bound shows that $z_{j}$ should be set inversely proportional
to $P_{j}$ to keep the rate loss constant. The splitting ratio $\bar{\rho}$
can be set to keep the constant loss within the specified limit. In
the following, SWIPT maximization problem is simplified. 

\subsection{Rate-energy maximization problem }

For the SWIPT precoding in literature \cite{6957150,Zhao2015}, authors
have formulated an optimization problem in which a linear sum of the
sum rate and sum harvested energy is maximized subjected to the quality-of-service
(QoS) constraints and the precoder constraints as 
\begin{subequations}
\begin{gather}
\max_{\mathbf{V}_{j},\forall j}\sum_{k}R_{k}\left(\mathbf{V}_{j},\forall j\big|\mathbf{H}_{j}\right)+\nu Q_{r}\left(\mathbf{V}_{j},\forall j\big|\mathbf{H}_{j}\right)\\
\text{subject to }\left|R_{j}-\bar{R}_{j}\right|\leq\frac{d}{2}\log_{2}c,\|\mathbf{V}_{j}\|_{F}^{2}\leq d,\forall j,
\end{gather}
\end{subequations}
 where $\nu$ is the weight controlling the preferred objective, and
$\bar{R}_{k}$ and $\frac{d}{2}\log_{2}c$ are the QoS rate constraint
and the specified rate-loss upper bound for the $j^{th}$ user. Note
that the above two are opposing objectives, i.e., if the sum rate
is maximized, the harvested energy is reduced, and if the sum harvested
energy is maximized, the sum rate degrades. %
To provide a balanced precoder, we start with the sum rate optimal
precoder, i.e., the TWRIA precoder $\mathbf{V}_{j}$, and degrade
this precoder in such a way that the degraded precoder satisfies the
required QoS constraint. In general, if we degrade the TWRIA precoder,
it will result in severe rate loss, causing the unexpected loss in
degrees of freedom. Thus, to avoid unexpected losses, we employ the
CD decomposition, in which the value of CD decides the degradation
in the precoder, i.e., the losses in degrees of freedom. It can be
seen from Lemma \ref{lem:Rate-loss-basic} that if the CD value is
chosen inversely proportional to SNR, there is no-loss of DoFs, that
is, only a constant rate loss is present. This constant rate loss
can be reduced via the splitting ratio. 

For example, from Lemma \ref{lem:Rate-loss-basic}, to keep the rate
loss upper bound to be a constant (say $\frac{d}{2}\log_{2}c$), the
required values of CD and splitting ratio can be computed via the
roots of the following equation 
\begin{align*}
 & \frac{M_{d}\bar{\rho}P_{j}z_{j}\beta_{rj}}{\bar{\rho}\left(\sum_{j=1}^{2K}P_{j}\frac{\beta_{rj}\sigma^{2}}{\beta_{kr}P_{r}}+\bar{\sigma}_{kr}^{2}\right)+\bar{\delta}_{kr}^{2}}\leq\frac{c-1}{2(K-1)}
\end{align*}
which can be rearranged into $z_{j}\leq\min\left(\bar{z}_{j}(\bar{\rho},c),z_{j}^{EH},\forall j\right)$
with 
\begin{align*}
 & \bar{z}_{j}(\bar{\rho},c)=\frac{(c-1)}{2P_{j}\beta_{rj}M_{d}(K-1)}\left[\sum_{j=1}^{2K}P_{j}\frac{\beta_{rj}\sigma^{2}}{\beta_{kr}P_{r}}+\bar{\sigma}_{R}^{2}+\frac{\bar{\delta}^{2}}{\bar{\rho}}\right].
\end{align*}
In the above, one need to have $\bar{z}_{j}(\bar{\rho},c)\leq z_{j}^{EH}$;
otherwise, the sum rates will be much worse due to the lack of interference
alignment, and in that scenario, EH maximized precoder would be the
better choice. In case of high SNR regime, these conditions can easily
met, since $z_{j}$ is inversely proportional to $P_{j}$. In case
of low and mid-SNR range, the value of $\bar{\rho}$ can be finely
tuned to get $z_{j}$ value under the limit. Therefore, given the
CD values and the IA precoders for the specified constant rate loss
upper bound, the corresponding balanced precoders are obtained in
the following section. 

\section{Proposed balanced precoding method\label{iv}}

\subsection{Optimization Problem}

Given the TWRIA precoders $\left\{ \mathbf{V}_{j},\forall j\right\} $
and the value of CD $\left\{ z_{j},\forall j\right\} $, we can now
focus on maximizing the harvested energy, since the expected sum rate
losses obtained with a given CD have a fixed and known upper bound.
Thus, the $j^{th}$ balanced precoder can be expressed using CD decomposition
from the Corollary \ref{cor:Given-CD} as 
\begin{equation}
\mathbf{V}_{j}^{BAL}=\mathbf{V}_{j}\mathbf{X}_{j}\mathbf{Y}_{j}+\mathbf{V}_{j}^{\text{null}}\mathbf{S}_{j}\mathbf{Z}_{j},
\end{equation}
where $\mathbf{Y}_{j}$ and $\mathbf{Z}_{j}$ are diagonal matrices;
the matrices $\mathbf{X}_{j}$ and $\mathbf{Z}_{j}$ are obtained
in the following to maximize the energy; and $\mathbf{V}_{j}^{\text{null}}$
represents the left null space of $\mathbf{V}_{j}$, i.e., $\mathbf{V}_{j}^{\text{null}}=\text{null}(\mathbf{V}_{j})\in\mathcal{G}_{M,M-d}$
such that $\mathbf{V}_{j}^{H}\mathbf{V}_{j}^{\text{null}}=\mathbf{0}$.

The optimization problem to find the balanced precoding to maximize
the total harvested energy can be cast for each $j^{th}$ precoder
as 
\begin{subequations}
\begin{align}
\max_{\mathbf{S}_{j},\mathbf{Z}_{j},\mathbf{X}_{j},\mathbf{Y}_{j}}\left\Vert \mathbf{H}_{rj}\mathbf{V}_{j}^{BAL}\right\Vert _{F}^{2}\label{eq:porb_state2}\\
\text{subject to }\mathbf{V}_{j}^{BAL}=\mathbf{V}_{j}\mathbf{X}_{j}\mathbf{Y}_{j}+\mathbf{V}_{j}^{\text{null}}\mathbf{S}_{j}\mathbf{Z}_{j},\label{eq:VBALCON}\\
\text{tr}\left(\mathbf{Z}_{j}\mathbf{Z}^{H}\right)=\text{tr}\left(\mathbf{I}-\mathbf{Y}_{j}\mathbf{Y}_{j}^{H}\right)\leq z_{j},\label{eq:z_j_con}\\
\mathbf{Z}_{j},\mathbf{Y}_{j}\text{ are diagonal matrices},\\
\mathbf{X}_{j}^{H}\mathbf{X}_{j}=\mathbf{X}_{j}\mathbf{X}_{j}^{H}=\mathbf{I}_{d},\label{eq:Xj_con}\\
\mathbf{S}_{j}^{H}\mathbf{S}_{j}=\mathbf{I}_{d}.
\end{align}
\end{subequations}
 The solution to the above problem is obtained as follows. First,
$\mathbf{S}_{j}$ is computed, followed by the computation of $\mathbf{Z}_{j}$
and $\mathbf{X}_{j}$.

\subsection{Getting $\mathbf{S}_{j}$}

Using the triangle inequality, the objective function in \eqref{eq:porb_state2}
can be upper bounded as 
\begin{align}
 & \left\Vert \mathbf{H}_{rj}\left(\mathbf{V}_{j}\mathbf{X}_{j}\mathbf{Y}_{j}+\mathbf{V}_{j}^{\text{null}}\mathbf{S}_{j}\mathbf{Z}_{j}\right)\right\Vert _{F}\nonumber \\
 & \leq\left\Vert \mathbf{H}_{rj}\mathbf{V}_{j}\mathbf{X}_{j}\mathbf{Y}_{j}\right\Vert _{F}+\left\Vert \mathbf{H}_{rj}\mathbf{V}_{j}^{\text{null}}\mathbf{S}_{j}\mathbf{Z}_{j}\right\Vert _{F},\label{eq:HVABL_tri}
\end{align}
where the equality occurs when both $\mathbf{H}_{rj}\mathbf{V}_{j}\mathbf{X}_{j}\mathbf{Y}_{j}$
and $\mathbf{H}_{rj}\mathbf{V}_{j}^{\text{null}}\mathbf{S}_{j}\mathbf{Z}_{j}$
are in the same direction or proportional to each other. Since both
the precoder $\mathbf{V}_{j}$ and its null space $\mathbf{V}_{j}^{\text{null}}$
are present in the above norm expression, the equality cannot be achieved
when $z_{j}>0$ or $\mathbf{Z}_{j}\neq\mathbf{0}$. Best efforts can
be done to align these matrices using the following optimization problem
as 
\begin{gather}
\arg\min_{\mathbf{S}_{j},\mathbf{Z}_{j},\mathbf{X}_{j},\mathbf{Y}_{j}}d_{c}^{2}\left(\mathbb{O}\left[\mathbf{H}_{rj}\mathbf{V}_{j}\mathbf{X}_{j}\mathbf{Y}_{j}\right],\mathbb{O}\left[\mathbf{H}_{rj}\mathbf{V}_{j}^{\text{null}}\mathbf{S}_{j}\mathbf{Z}_{j}\right)\right],\nonumber \\
\stackrel{(a)}{=}\arg\min_{\mathbf{S}_{j}}d_{c}^{2}\left(\mathbb{O}\left[\mathbf{H}_{rj}\mathbf{V}_{j}\right],\mathbb{O}\left[\mathbf{H}_{rj}\mathbf{V}_{j}^{\text{null}}\mathbf{S}_{j}\right)\right],\nonumber \\
\stackrel{(b)}{=}\arg\max_{\mathbf{S}_{j}^{H}\mathbf{S}_{j}=\mathbf{I}}\text{tr}\left(\mathbf{D}_{Vj}\mathbf{V}_{j}^{H}\mathbf{H}_{rj}^{H}\mathbf{H}_{rj}\mathbf{V}_{j}^{\text{null}}\mathbf{S}_{j}\mathbf{D}_{Vnj}\right),
\end{gather}
where in $(a)$, the orthogonalization property is used, since both
matrices represent the same basis of the column space; in \textbf{$(b)$},
the definition of CD, $\mathbb{O}\left[\mathbf{A}\right]=\mathbf{A}\left(\mathbf{A}^{H}\mathbf{A}\right)^{-1/2}$,
$\mathbf{D}_{Vj}=\left(\mathbf{V}_{j}^{H}\mathbf{H}_{rj}^{H}\mathbf{H}_{rj}\mathbf{V}_{j}\right)^{-1/2}$,
and $\mathbf{D}_{Vnj}=\left(\mathbf{S}_{j}^{H}\mathbf{V}_{j}^{\text{null}H}\mathbf{H}_{rj}^{H}\mathbf{H}_{rj}\mathbf{V}_{j}^{\text{null}}\mathbf{S}_{j}\right)^{-1/2}$
are used. From $(b)$, the solution is obtained by choosing the columns
in the same directions as $\mathbf{V}_{j}^{\text{null}H}\mathbf{H}_{rj}^{H}\mathbf{H}_{rj}\mathbf{V}_{j}$
to maximize the trace-value as 
\begin{align}
\mathbf{S}_{j} & =\mathbb{O}\left[\mathbf{V}_{j}^{\text{null}H}\mathbf{H}_{rj}^{H}\mathbf{H}_{rj}\mathbf{V}_{j}\mathbf{D}_{Vj}\mathbf{D}_{Vnj}\right]\nonumber \\
 & \equiv\mathbb{O}\left[\mathbf{V}_{j}^{\text{null}H}\mathbf{H}_{rj}^{H}\mathbf{H}_{rj}\mathbf{V}_{j}\right],
\end{align}
where the equivalence can be considered due to the fact that $\mathbf{X}_{j}$,
$\mathbf{Y}_{j}$ and $\mathbf{Z}_{j}$ are unknown, and thus, $\mathbf{S}_{j}$
can be independently and equivalently computed first. Further, letting
$\mathbf{A}_{j}=\mathbf{V}_{j}^{\text{null}H}\mathbf{H}_{j}^{H}\mathbf{H}_{j}\mathbf{V}_{j}$,
the cross-term below can be simplified as $\text{tr}\left(\mathbf{Y}_{j}^{H}\mathbf{X}_{j}^{H}\mathbf{V}_{j}^{H}\mathbf{H}_{j}^{H}\mathbf{H}_{j}\mathbf{V}_{j}^{\text{null}}\mathbf{S}_{j}\mathbf{Z}_{j}\right)$$=\text{tr}\left(\mathbf{Z}_{j}\mathbf{Y}_{j}^{H}\mathbf{X}_{j}^{H}\left(\mathbf{A}_{j}^{H}\mathbf{A}_{j}\right)^{1/2}\right).$

\subsection{Getting $\mathbf{Z}_{j}$ and $\mathbf{X}_{j}$: an iterative approach}

Further, from \eqref{eq:HVABL_tri}, squaring the terms on both sides
yields the Cauchy Schwarz's inequality 
\begin{align}
 & \Re\left[\text{tr}\left(\mathbf{Y}_{j}^{H}\mathbf{X}_{j}^{H}\mathbf{V}_{j}^{H}\mathbf{H}_{j}^{H}\mathbf{H}_{j}\mathbf{V}_{j}^{\text{null}}\mathbf{S}_{j}\mathbf{Z}_{j}\right)\right]\nonumber \\
 & \leq\left\Vert \mathbf{H}_{j}\mathbf{V}_{j}\mathbf{X}_{j}\mathbf{Y}_{j}\right\Vert _{F}\left\Vert \mathbf{H}_{j}\mathbf{V}_{j}^{\text{null}}\mathbf{S}_{j}\mathbf{Z}_{j}\right\Vert _{F},
\end{align}
which suggests that equivalently, the above cross-term can be maximized
to get the maximum harvested energy.

Since the matrices $\mathbf{Y}_{j}$ and $\mathbf{Z}_{j}$ are diagonal,
the matrix $\mathbf{Y}_{j}=\mathcal{D}\left(y_{j1},\ldots,y_{jd}\right)$
can be obtained from $\mathbf{Z}_{j}=\mathcal{D}\left(z_{j1},\ldots,z_{jd}\right)$
using the constraint in \eqref{eq:z_j_con} and \eqref{eq:decomp-YY-1}
as 
\begin{equation}
y_{ji}=+\sqrt{1-z_{ji}^{2}},\forall i=1,\ldots,d,\label{eq:Y_chol1}
\end{equation}
satisfying the constraint in \eqref{eq:z_j_con}. The remaining components
of the CD decomposition can be computed as the solution to the following
optimization problem as 
\begin{align*}
 & \max_{\mathbf{Z}_{j},\mathbf{X}_{j}}\Re\left[\text{tr}\left(\mathbf{Y}_{j}^{H}\mathbf{X}_{j}^{H}\mathbf{V}_{j}^{H}\mathbf{H}_{rj}^{H}\mathbf{H}_{rj}\mathbf{V}_{j}^{\text{null}}\mathbf{S}_{j}\mathbf{Z}_{j}\right)\right],
\end{align*}
which is a non-convex problem due to the product of $\mathbf{Z}_{j}$
and $\mathbf{X}_{j}$. The efficient way to solve the problem is via
an iterative method, where $\mathbf{X}_{j}$ and $\mathbf{Z}_{j}$
are solved alternately. Given $\mathbf{Z}_{j}$ and $\mathbf{Y}_{j}$,
the optimization problem above can be reduced to a convex problem
for $\mathbf{X}_{j}$ as 
\begin{subequations}
\begin{align}
 & \max_{\mathbf{X}_{j}}\Re\left[\text{tr}\left(\mathbf{Y}_{j}^{H}\mathbf{X}_{j}^{H}\mathbf{V}_{j}^{H}\mathbf{H}_{rj}^{H}\mathbf{H}_{rj}\mathbf{V}_{j}^{\text{null}}\mathbf{S}_{j}\mathbf{Z}_{j}\right)\right]\label{eq:optX}\\
 & \text{subject to }\|\mathbf{X}_{j}\|\leq1,
\end{align}
\end{subequations}
 where the spectral norm constraint above leads to the same constraint
in \eqref{eq:Xj_con}. The solution for $\mathbf{X}_{j}$ can be obtained
by choosing the same column directions as of $\mathbf{V}_{j}^{H}\mathbf{H}_{rj}^{H}\mathbf{H}_{rj}\mathbf{V}_{j}^{\text{null}}\mathbf{S}_{j}\mathbf{Z}_{j}\mathbf{Y}_{j}^{H}$,
i.e., 
\begin{align}
\mathbf{X}_{j} & =\mathbb{O}\left[\mathbf{V}_{j}^{H}\mathbf{H}_{rj}^{H}\mathbf{H}_{rj}\mathbf{V}_{j}^{\text{null}}\mathbf{S}_{j}\mathbf{Z}_{j}\mathbf{Y}_{j}^{H}\right].
\end{align}
Note that the above $\mathbf{X}_{j}$ cannot be equivalently set to
$\mathbb{O}\left[\mathbf{V}_{j}^{H}\mathbf{H}_{rj}^{H}\mathbf{H}_{rj}\mathbf{V}_{j}^{\text{null}}\mathbf{S}_{j}\right]$,
since the above particular directions are important. Further, substituting
$\mathbf{X}_{j}$ in the trace yields the following result. 
\begin{prop}
\label{prop:trace-val-noneg}With the above selection of $\mathbf{X}_{j}$,
the trace-value is non-negative 
\begin{align*}
 & \text{tr}\left(\mathbf{Y}_{j}^{H}\mathbf{X}_{j}^{H}\mathbf{V}_{j}^{H}\mathbf{H}_{rj}^{H}\mathbf{H}_{rj}\mathbf{V}_{j}^{\text{null}}\mathbf{S}_{j}\mathbf{Z}_{j}\right)=\text{tr}\left(\left(\mathbf{B}_{j}^{H}\mathbf{B}_{j}\right)^{1/2}\right)\geq0,
\end{align*}
where $\mathbf{B}_{j}=\mathbf{V}_{j}^{H}\mathbf{H}_{rj}^{H}\mathbf{H}_{rj}\mathbf{V}_{j}^{\text{null}}\mathbf{S}_{j}\mathbf{Z}_{j}\mathbf{Y}_{j}^{H}=\left(\mathbf{A}_{j}^{H}\mathbf{A}_{j}\right)^{1/2}$
$\mathbf{Z}_{j}\mathbf{Y}_{j}^{H}$, and the equality occurs when
$z_{j}=0$. 
\end{prop}
Next, given $\mathbf{Y}_{j}$, $\mathbf{X}_{j}$ and $z_{j}<z_{j}^{EH}$,
the diagonal matrix $\mathbf{Z}_{j}$ can be obtained from the following
convex problem as 
\begin{subequations}
\begin{align}
 & \max_{\mathbf{Z}_{j}}\Re\left[\text{tr}\left(\mathbf{Y}_{j}^{H}\mathbf{X}_{j}^{H}\mathbf{V}_{j}^{H}\mathbf{H}_{rj}^{H}\mathbf{H}_{rj}\mathbf{V}_{j}^{\text{null}}\mathbf{S}_{j}\mathbf{Z}_{j}\right)\right]\label{eq:optZ}\\
 & \text{subject to }\|\mathbf{Z}_{j}\|_{F}\leq\sqrt{z_{j}},\label{eq:Z_con1}\\
 & \mathbf{Z}_{j}\text{ is a diagonal matrix},\label{eq:Z_con2diag}\\
 & \mathbf{0}\preceq\mathbf{Z}_{j}\preceq\mathbf{I}_{d}.\label{eq:Z01con}
\end{align}
\end{subequations}
We can equivalently recast the above problem as 
\begin{subequations}
\begin{align}
 & \max_{z_{ji},,\forall i}\sum_{i=1}^{d}c_{ji}z_{ji}\\
 & \text{subject to }\sum_{i=1}^{d}z_{ji}^{2}\leq\sqrt{z_{j}},\\
 & 0\leq z_{ji}\leq1,\forall i=1,\ldots,d,
\end{align}
\end{subequations}
 where the vector $c_{ji}=\left[\mathbf{Y}_{j}^{H}\mathbf{X}_{j}^{H}\mathbf{V}_{j}^{H}\mathbf{H}_{rj}^{H}\mathbf{H}_{rj}\mathbf{V}_{j}^{\text{null}}\mathbf{S}_{j}\right]_{i,i},\forall i=1,\ldots,d$.
The values $c_{ji},\forall i$ are real and non-negative from the
proposition \ref{prop:trace-val-noneg}. The solution to the above
problem is given by choosing $\mathbf{z}_{j}$ equal to $\mathbf{c}_{j}$
and scaling it to satisfy the norm constraint. Thus, we write $z_{ji}=\min\left(\sqrt{z_{j}}\frac{c_{ji}}{\|\mathbf{c}_{j}\|},1\right)$,
and normalize the resulting entries to satisfy $\sum_{i\in\mathcal{I}}z_{ji}^{2}=z_{j}-\left(d-\left|\mathcal{I}\right|\right)$,
where $\mathcal{I}=\left\{ i:z_{ji}<1\right\} $, i.e., $z_{ji}\leftarrow\frac{z_{ji}}{\sum_{i\in\mathcal{I}}z_{ji}^{2}}\sqrt{z_{j}-\left(d-\left|\mathcal{I}\right|\right)},\forall i\in\mathcal{I}$.

\subsection{Iterative CD algorithm}

\begin{algorithm}
\begin{algorithmic}[1]

\Require{$\mathbf{H}_{rj}$, $\mathbf{V}_{j}$ and $z_{j}$.}

\Ensure{$\mathbf{V}_{j}^{BAL}$.}

\If{ $z_{j}>z_{j}^{EH}$}

\State{Return $\mathbf{V}_{j}^{BAL}=\mathbf{V}_{j}^{EH}$.}

\Else

\State{Compute $\mathbf{S}_{j}=\mathbb{O}\left[\mathbf{V}_{j}^{\text{null}H}\mathbf{H}_{rj}^{H}\mathbf{H}_{rj}\mathbf{V}_{j}\right]$.}

\State{Initialize $\mathbf{Z}_{j}=\sqrt{\frac{z_{j}}{d}}\mathbf{I}_{d}$
and $\mathbf{Y}_{j}$ by \eqref{eq:Y_chol1}.}

\State{\label{step}Solve \eqref{eq:optX} to get $\mathbf{X}_{j}$.}

\State{Solve \eqref{eq:optZ} to get $\mathbf{Z}_{j}$.}

\State{Get $\mathbf{Y}_{j}$ by \eqref{eq:Y_chol1}.}

\State{Go to step \ref{step} until convergence. }

\State{Return $\mathbf{V}_{j}^{BAL}$ via \eqref{eq:VBALCON}.}

\EndIf

\end{algorithmic}\caption{Iterative CD decomposition procedure.\label{alg:Iterative-CD-decomposition}}
\end{algorithm}

Now, with all components obtained, the resulting balanced precoder
can be computed via \eqref{eq:VBALCON}. The summary of this procedure
is given in Algorithm \ref{alg:Iterative-CD-decomposition}. If $z_{j}>z_{j}^{EH}$,
we choose the energy optimized precoder as the balanced precoder $\mathbf{V}_{j}^{BAL}=\mathbf{V}_{j}^{EH}$.
Regarding the convergence, it can be seen that since both $\mathbf{Z}_{j}$
and $\mathbf{X}_{j}$ maximize the same linear objective, thus convergence
is guaranteed with a global optimum value. Regarding the number of
iterations, we observe via simulations that it takes only a few ($4$
to $8$) iterations to converge.

\subsection{Computational complexity}

The product $\mathbf{H}_{j}^{H}\mathbf{H}_{j}$ and its EVD need $\mathcal{O}\left(M^{2}R\right)$
and $\mathcal{O}\left(M^{3}\right)$ operations. For $\mathbf{S}_{j}$,
the product and $\mathbb{O}[\cdot]$ need $\mathcal{O}\left(M^{2}R\right)$
and $\mathcal{O}\left(d^{2}\cdot(M-d)+d^{3}\right)=\mathcal{O}\left(Md^{2}\right)$
operations. The rest of operations are below $\mathcal{O}\left(M^{3}\right)$,
since $R$ is the order of $M$. Thus, Algorithm \ref{alg:Iterative-CD-decomposition}
has $\mathcal{O}\left(M^{3}+Md^{2}N_{I}\right)\approx\mathcal{O}\left(M^{3}\right)$
computational complexity, where the number of iterations $N_{I}$
for convergence are few ($4$ to $8$), i.e., $N_{I}\ll\frac{M^{2}}{d^{2}}$.

\subsection{Bounds}

Note that any trivial balanced precoding cannot guarantee better harvested
energy. Thus, for the proposed balanced precoding, the following bounds
can be obtained. 
\begin{lem}
\label{lem:bounds}Given the balanced precoding $\left\{ \mathbf{V}_{k}^{BAL},\forall k\right\} $
for the channel $\left\{ \mathbf{H}_{kj},\forall k,j\right\} $ with
the TWRIA precoders $\left\{ \mathbf{V}_{k},\forall k\right\} $,
the total harvested energy can be bounded as 
\begin{align}
\zeta\rho\sum_{j=1}^{2K}\frac{P_{j}}{d} & \left[\left\Vert \mathbf{H}_{rj}\mathbf{V}_{j}\right\Vert _{F}^{2}\left(1-\frac{z_{j}}{d}\right)+\left\Vert \mathbf{H}_{rj}\mathbf{V}_{j}^{\text{n}}\right\Vert _{F}^{2}\left(\frac{z_{j}}{d}\right)\right]\nonumber \\
 & \leq Q_{r}(\rho,\mathbf{V}_{j}^{BAL})\leq\zeta\rho\sum_{j=1}^{2K}P_{j}\lambda_{j1}.
\end{align}
\end{lem}
\begin{IEEEproof}
Proof is given in Appendix\ref{subsec:Proof-of-bounds}. 
\end{IEEEproof}
The above result shows an improvement over \eqref{eq:max_EH-Qk},
i.e., the balanced precoding promises a better harvested energy than
that achieved using just the perfect IA precoders, if $z_{j}>0$ for
any $j$, and $\rho>0$. The corresponding resulting rate loss can
be obtained from the upper bound in the Lemma \ref{lem:Rate-loss-basic}.

With $\mathcal{CN}(0,1)$ entries for the matrix $\mathbf{H}_{kj}$
and $P_{j}=P,\forall j$, performing the expectation on both sides
in the above equation gives 
\begin{equation}
2KP\zeta\rho R\approx\mathbb{E}\left\{ Q_{r}(\rho)\right\} \leq2KP\zeta\rho Rd\left(\frac{R+d}{Rd+1}\right)^{2/3},
\end{equation}
where the left approximation is obtained assuming $\mathbb{E}\left\{ \left\Vert \mathbf{H}_{rj}\mathbf{V}_{j}^{BAL}\right\Vert _{F}^{2}\right\} $
$\approx$ $Rd$, and the right inequality is given by $\mathbb{E}\left\{ \lambda_{j1}\right\} =Rd\left(\frac{R+d}{Rd+1}\right)^{2/3}$
\cite{5770238}.

\section{Simulation Results\label{v}}

We consider $2K=6$ nodes, each transmitting $d=2$ data streams via
a relay having antennas $R=Kd=M=6$, i.e., $\left(6,6,2\right)^{6}$
system, which is analogous to an IC system $\left(6\times6,2\right)^{6}$.
The noise variances at receivers is assumed to be unity, $\sigma^{2}=\sigma_{R}^{2}=1$,
while the noise variance at the splitters is assumed to be $\delta^{2}=0.1$,
when $\rho>0$; the EH conversion efficiency is set to be $\zeta=0.5$;
and the transmit power at the relay is considered same as the transmit
power of users $P_{j}=P_{r},\forall j$. For the balanced precoding,
the iterative CD algorithm is run for $6$ iterations. QPSK symbol
error rate performance is averaged over $20,000$ symbols. In the
following figures, we compare three different precoding strategies
given below. 
\begin{itemize}
\item (MAX-EH) Harvested energy maximizing precoder; 
\item (Span-IA) Balanced precoders from subspace alignment method with $z=0,0.1$
\cite{6516875,5956515}; 
\item (TWRIA) Balanced precoder from MMSE based IA algorithm \cite{9097459}
with $z=0,0.1$.
\end{itemize}

\subsection{Convergence of TWRIA Algorithm }

\begin{figure}
\centering\includegraphics[width=1\columnwidth]{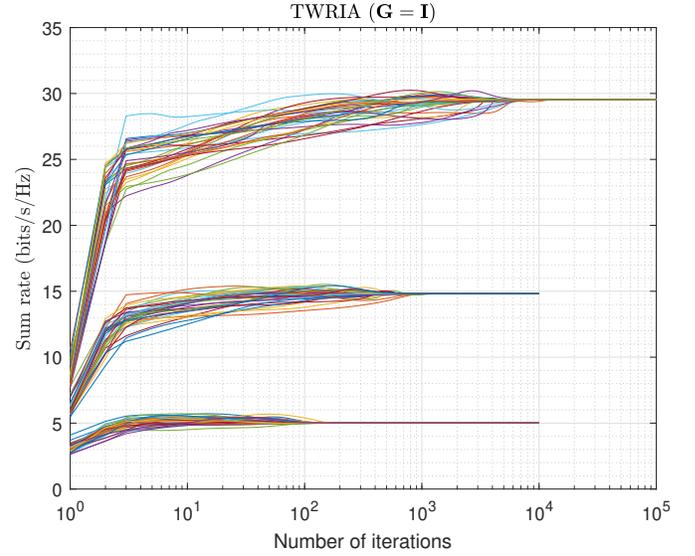}\caption{Convergence of TWRIA algorithm with $32$ different initializations
for $\left(6\times6,2\right)^{6}$ TWR system for $8.5$, $17$ and
$25$ dB SNR values, respectively. \label{fig:Convergence-of-two-way}}
\end{figure}
Figure \ref{fig:Convergence-of-two-way} illustrates the sum rate
of a $(6\times6,2)^{6}$ system versus the number of iterations for
$32$ different precoder initializations for TWRIA Algorithm \ref{alg:2WRIA_algo}
with respect to different values of SNR. As the iteration number increases,
the sum rate improves and converges globally after enough iterations.
The rate of convergence, i.e., the number of required iterations for
convergence, depends on the operating SNR. An average of $10^{3}$,
$10^{4}$ and $10^{5}$ iterations are required for convergence for
$8.5$, $17$ and $25$ dB SNR values, respectively.

\subsection{Sum rate versus SNR }

\begin{figure}
\centering\includegraphics[width=1\columnwidth]{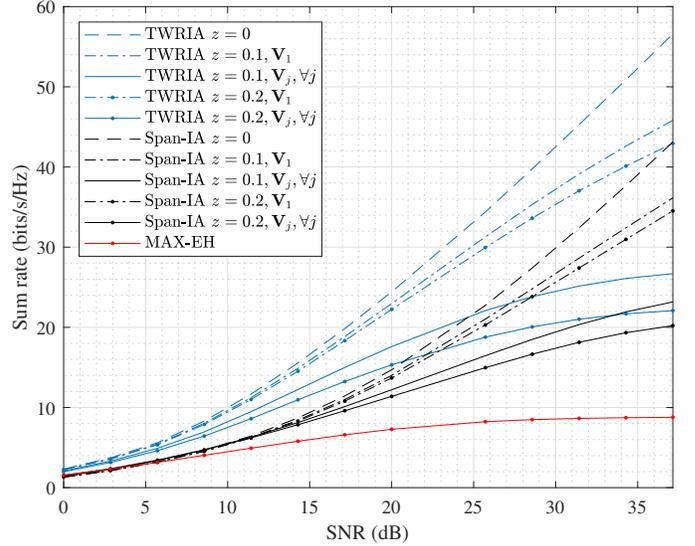}\caption{Sum rate versus SNR plot for TWR system for $\left(6\times6,2\right)^{6}$
system. \label{fig:sr_vs_SNR}}
\end{figure}
Figure \ref{fig:sr_vs_SNR} plots the sum rate versus SNR values for
three types of precoders. First, it can be seen that the TWRIA algorithm
provides better sum rates, which scales linearly with SNRs, as compared
with the subspace alignment method (span-IA). Also, when the precoder
is balanced for better energy harvesting, sum rates are decreased.
When only single user employs balanced precoding, the sum rate is
not reduced by a large amount, i.e., there is much less degree of
freedom loss, as compared to the case when all users use the balanced
precoding with the same CD value. Based on the required energy at
the relay, the precoding can be balanced at users via different CD
values. For a fixed CD value, the corresponding rate loss show an
increase with SNR, as analyzed earlier in Lemma \ref{lem:Rate-loss-basic}.
Further, the decrements of sum rate with respect to the CD values
can also be seen in the following rate-energy plots.

\subsection{Rate-energy plots}

Given the precoders $\left\{ \mathbf{V}_{k},\forall k\right\} $,
the rate-energy region can be written as 
\begin{equation}
\mathcal{C}=\left\{ \left(R,Q\right):R\leq\sum_{k=1}^{K}R_{k}\left(\bar{\rho},\mathbf{V}_{k}\right),Q\leq Q_{r}\left(\rho,\mathbf{V}_{k}\right)\right\} .
\end{equation}
\begin{figure}
\centering\includegraphics[width=1\columnwidth]{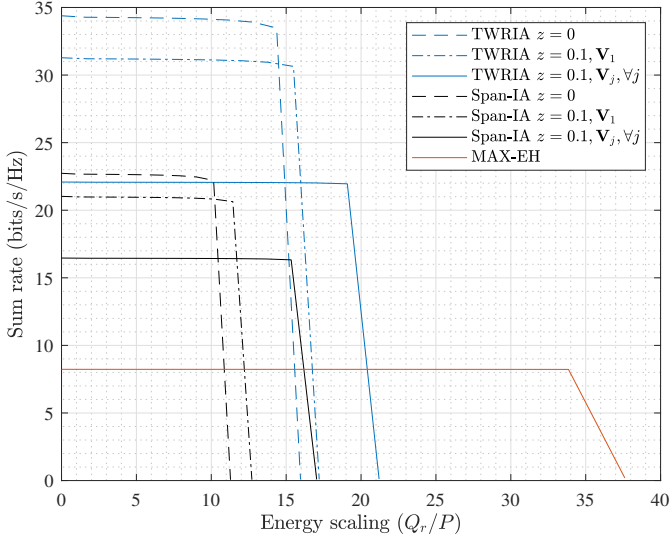}\caption{Rate-energy plots for TWR $\left(6\times6,2\right)^{6}$ system for
different methods at $25$ dB SNR. \label{fig:re_plot_all}}
\end{figure}
For a splitting noise variance $\delta^{2}=0.1$, the parametric plots
are drawn to illustrate rate-energy regions \cite{6623062,8241822}.
Figure \ref{fig:re_plot_all} compares the same rate-energy region
for different precoding schemes. The trend of different methods with
various CD values is similar as in Figure \ref{fig:sr_vs_SNR}. Different
rates and energy plots conclude that energy harvesters can be improved
at the expense of sum rate optimality.

\begin{figure}
\centering\includegraphics[width=1\columnwidth]{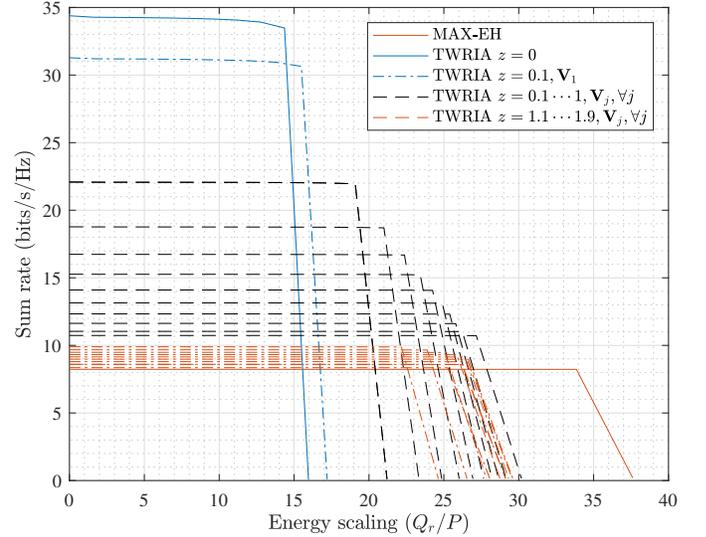}\caption{Rate-energy plots for TWRIA algorithm for TWR $\left(6\times6,2\right)^{6}$
system at $25$ dB SNR. \label{fig:re_plot_IA}}
\end{figure}

Figure \ref{fig:re_plot_IA} shows the sum rate versus the harvested
energy plot for aforementioned precoders with and without balanced
precoding. It can be noted that the TWRIA region provides higher sum
rates and lower energies, while the region for MAX-EH precoders has
less sum rates and higher energies. These plots represent two extreme
ends of rate and energy achievability. Next, for the balanced precoding,
it can be observed that as $z$ increases, the rate decreases and
the energy increases, when $z<\min_{j}z_{j}^{EH}$, where $\min_{j}z_{j}^{EH}$
represents the threshold for CD. When $z>\min_{j}z_{j}^{EH}$, both
rate and energy achieved are lower. Therefore, the value of CD ($z$)
must be properly selected below the threshold to balance both the
rates and the energy at each user.

\begin{figure}
\centering\includegraphics[width=1\columnwidth]{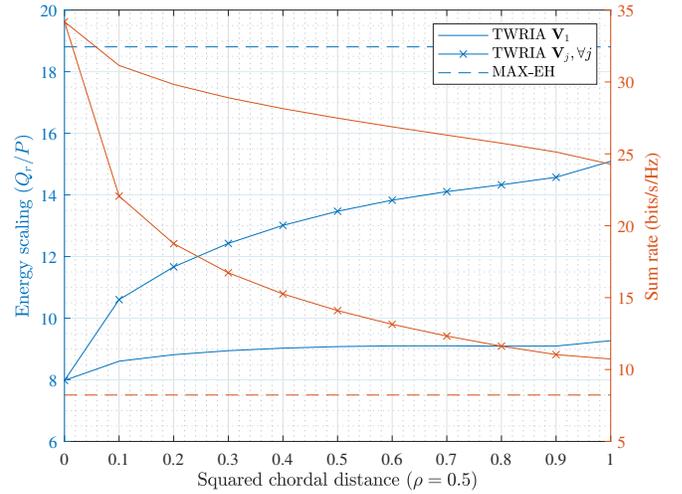}\caption{Rate-energy versus the squared CD plot for TWR system for $\left(6\times6,2\right)^{6}$
system at $25$ dB SNR. \label{fig:re_z}}
\end{figure}
Figure \ref{fig:re_z} plots the sum rate (right-axis in red) and
the harvested energy (left-axis in blue) versus the squared CD $z=d_{c}^{2}\left(\mathbf{V}_{j},\mathbf{V}_{j}^{BAL}\right),\forall j$
required for the balanced precoding with TWRIA and span-IA methods.
It can be seen that the sum rate decreases in a logarithmic manner
as $z$ increases. This behavior has been analyzed in the Lemma \ref{lem:Rate-loss-basic}
for the rate-loss upper bound. On the other hand, the harvested energy
increases if $z$ increases. %

\subsection{QPSK symbol error rate versus SNR}

\begin{figure}
\centering\includegraphics[width=1\columnwidth]{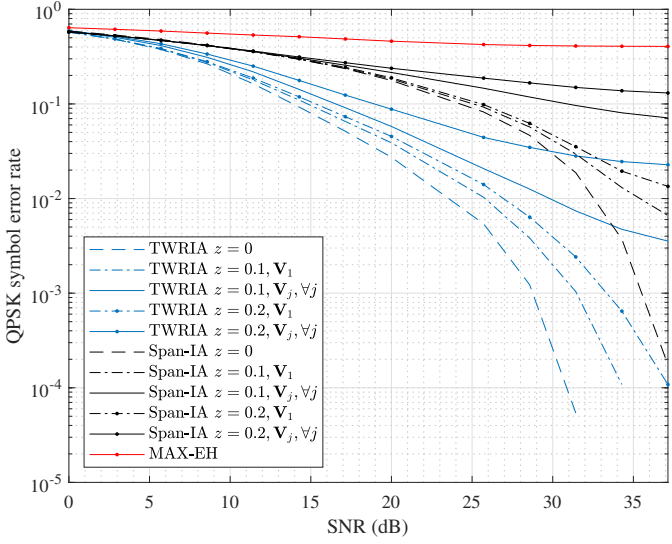}\caption{QPSK symbol error rate versus SNR plot for TWR system for $\left(6\times6,2\right)^{6}$
system. \label{fig:ser_vs_SNR}}
\end{figure}

Figure \ref{fig:ser_vs_SNR} depicts the average symbol error rate
(SER) plots with uncoded QPSK modulation for $(6\times6,2)^{6}$.
It can be seen that the perfect IA precoders ($z=0$) achieve the
minimum SER, while with $z=0.1$, the SER saturates. It can be noted
the trend of error rate curves is same as in Figure \ref{fig:sr_vs_SNR},
and the MAX-EH and the span-IA provide the worse SERs than that of
TWRIA.

\section{Conclusion\label{vi}}

In this paper, for the TWR system, a modified IA algorithm is presented
while considering AF relaying. To improve the SWIPT, i.e., to obtain
the better rate-energy trade-offs, the CD decomposition-based balance
precoding has been investigated, which is of low-complexity than the
other SDR based convex relaxation-based suboptimal methods in the
literature. Further, maximum energy based precoders, rate-loss upper
bound, and harvested energy bounds are derived to show and compare
the rate-energy trade-offs. Simulations with other IA methods conclude
the effectiveness of the proposed IA and the CD decomposition-based
balanced precoding schemes.

The future work is to investigate the effect of quantized or analog
feedback including the specific details of 5G New Radio scenarios
including imperfect CSI.

\appendices{}

\section*{Appendices}

\subsection{IA Algorithm for TWR system\label{APPIA}}

In the following, precoder and combiner expressions are derived to
minimize the mean squared error (MSE), followed by an iterative procedure.

\subsubsection{Receiver design}

Given the precoders $\mathbf{V}_{j}$ for $\forall j\neq k$, the
receive combiner $\mathbf{U}_{k}$ at the $k^{th}$ receiver can be
obtained to minimize the MSE \cite{9097459} as 
\begin{align*}
 & \min_{\mathbf{U}_{k}^{H}\mathbf{U}_{k}=\mathbf{I}_{d}}\mathbb{E}\left\Vert \hat{\mathbf{y}}_{k}-\mathbf{s}_{k'}\right\Vert _{2}^{2}.
\end{align*}
The MSE at the $k^{th}$ user can be simplified as 
\begin{align*}
 & \mathbb{E}\left\Vert \hat{\mathbf{y}}_{k}-\mathbf{s}_{k'}\right\Vert _{2}^{2}\\
 & =\mathbb{E}\|\left(\sqrt{\bar{\rho}}\mathbf{U}_{k}^{H}\mathbf{H}_{kk'}\mathbf{V}_{k'}-\mathbf{I}\right)\mathbf{s}_{k'}\|_{2}^{2}+\mathbb{E}\|\mathbf{U}_{k}^{H}\mathbf{n}_{k}\|_{2}^{2}\\
 & +\bar{\rho}\sum_{j\neq k,k'}\mathbb{E}\|\mathbf{U}_{k}^{H}\mathbf{H}_{kj}\mathbf{V}_{j}\mathbf{s}_{j}\|_{2}^{2}+\bar{\rho}\mathbb{E}\|\mathbf{U}_{k}^{H}\mathbf{H}_{kr}\mathbf{G}\tilde{\mathbf{w}}_{r}\|_{2}^{2}\\
 & =\|\sqrt{\bar{\rho}}\mathbf{U}_{k}^{H}\mathbf{H}_{kk'}\mathbf{V}_{k'}-\mathbf{I}\|_{F}^{2}\frac{P_{k'}}{d}+\sigma^{2}\|\mathbf{U}_{k}\|_{F}^{2}\\
 & +\bar{\rho}\sum_{j\neq k,k'}\frac{P_{j}}{d}\|\mathbf{U}_{k}^{H}\mathbf{H}_{kj}\mathbf{V}_{j}\|_{F}^{2}+\bar{\rho}\sigma_{ID}^{2}\|\mathbf{U}_{k}^{H}\mathbf{H}_{kr}\mathbf{G}\|_{F}^{2}\\
 & =P_{k'}-\frac{2\bar{\rho}P_{k'}}{d}tr\Re\left(\mathbf{U}_{k}^{H}\mathbf{H}_{kk'}\mathbf{V}_{k'}\right)\\
 & +\bar{\rho}\sum_{j\neq k}\frac{P_{j}}{d}\|\mathbf{U}_{k}^{H}\mathbf{H}_{kj}\mathbf{V}_{j}\|_{F}^{2}+tr\left(\mathbf{U}_{k}^{H}\mathbf{C}_{k}\mathbf{U}_{k}\right).
\end{align*}

Differentiating with respect to $\mathbf{U}_{k}$ gives 
\begin{align*}
 & \frac{\bar{\rho}P_{k'}}{d}\mathbf{H}_{kk'}\mathbf{V}_{k'}=\sum_{j\neq k}\frac{\bar{\rho}P_{j}}{d}\mathbf{H}_{kj}\mathbf{V}_{j}\mathbf{V}_{j}^{H}\mathbf{H}_{kj}^{H}\mathbf{U}_{k}+\mathbf{C}_{k}\mathbf{U}_{k},
\end{align*}
yielding $\mathbf{U}_{k}=$
\begin{equation}
\mathbb{O}\left[\left(\sum_{j\neq k}\frac{\bar{\rho}P_{j}}{d}\mathbf{H}_{kj}\mathbf{V}_{j}\mathbf{V}_{j}^{H}\mathbf{H}_{kj}^{H}+\mathbf{C}_{k}\right)^{-1}\mathbf{H}_{kk'}\mathbf{V}_{k'}\frac{\bar{\rho}P_{k'}}{d}\right],\label{eq:Uk}
\end{equation}
where $\mathbb{O}\left(\cdot\right)$ denotes the orthonormality operation
as defined in the notations section. 

\subsubsection{Precoder design}

Similarly, given the combiners $\mathbf{U}_{k}$, the transmit precoder
can be optimized to minimize the total MSE as 
\begin{align*}
 & \min_{\mathbf{V}_{j}^{H}\mathbf{V}_{j}=\mathbf{I}_{d}}\sum_{k}\mathbb{E}\left\Vert \hat{\mathbf{y}}_{k}^{H}-\mathbf{s}_{k'}^{H}\right\Vert _{2}^{2}.
\end{align*}
The total MSE across all receivers can be rearranged as 
\begin{align*}
 & \sum_{k}\mathbb{E}\left\Vert \hat{\mathbf{y}}_{k}^{H}-\mathbf{s}_{k'}^{H}\right\Vert _{2}^{2}\\
 & =\sum_{k'}P_{k'}-\sum_{k}\frac{2\bar{\rho}P_{k'}}{d}tr\Re\left(\mathbf{U}_{k}^{H}\mathbf{H}_{kk'}\mathbf{V}_{k'}\right)\\
 & +\bar{\rho}\sum_{k}\sum_{j\neq k}\frac{P_{j}}{d}\|\mathbf{U}_{k}^{H}\mathbf{H}_{kj}\mathbf{V}_{j}\|_{F}^{2}+\sum_{k}tr\left(\mathbf{U}_{k}^{H}\mathbf{C}_{k}\mathbf{U}_{k}\right)\\
 & =\bar{\rho}\sum_{j}\frac{P_{j}}{d}\sum_{k\neq j}\|\mathbf{V}_{j}^{H}\mathbf{H}_{kj}^{H}\mathbf{U}_{k}\|_{F}^{2}-\sum_{j}\frac{2\bar{\rho}P_{j}}{d}tr\Re\left(\mathbf{V}_{j}^{H}\mathbf{H}_{j'j}^{H}\mathbf{U}_{j}\right)\\
 & +\sum_{k'}P_{k'}+\sum_{k}tr\left(\mathbf{U}_{k}^{H}\mathbf{C}_{k}\mathbf{U}_{k}\right).
\end{align*}
Differentiating the above MSE with respect to $\mathbf{V}_{j}$ provides
\begin{align*}
 & \sum_{k\neq j}\mathbf{H}_{kj}^{H}\mathbf{U}_{k}\mathbf{U}_{k}^{H}\mathbf{H}_{kj}\mathbf{V}_{j}=\mathbf{H}_{j'j}^{H}\mathbf{U}_{j},
\end{align*}
and simplifying to 
\begin{equation}
\mathbf{V}_{j}=\mathbb{O}\left[\left(\sum_{k\neq j}\mathbf{H}_{kj}^{H}\mathbf{U}_{k}\mathbf{U}_{k}^{H}\mathbf{H}_{kj}+\epsilon\mathbf{I}\right)^{-1}\mathbf{H}_{j'j}^{H}\mathbf{U}_{j}\right],\label{eq:Vk}
\end{equation}
where $\epsilon\mathbf{I}_{M}$ is added as a regularization term
to avoid singular matrix for inverse. The value of regularizer parameter
is chosen $\frac{d}{P_{j}}\cdot\frac{tr(\mathbf{C}_{k})}{M}$, that
is close to the normalized noise power at the receiver $j$. %

\subsubsection{Relay design (amplify-and-forward)}

For an amplified and forward relay with $\mathbf{G}=\alpha\mathbf{I}$,
the factor $\alpha$ can be obtained by solving the power constraint
$tr\mathbb{E}\mathbf{G}\mathbf{y}_{r}^{ID}\mathbf{y}_{r}^{IDH}\mathbf{G}^{H}=\text{tr}\left(\mathbf{G}\left(\sum_{j=1}^{2K}\mathbf{H}_{rj}\mathbf{V}_{j}\mathbf{V}_{j}^{H}\mathbf{H}_{rj}^{H}\bar{\rho}\frac{P_{j}}{d}+\bar{\rho}\sigma_{ID}^{2}\mathbf{I}_{R}\right)\mathbf{G}^{H}\right)=P_{r}$.
Thus, we have
\begin{equation}
\alpha^{2}=\frac{P_{r}}{\text{tr}\left(\sum_{j=1}^{2K}\bar{\rho}\frac{P_{j}}{d}\mathbf{H}_{rj}\mathbf{V}_{j}\mathbf{V}_{j}^{H}\mathbf{H}_{rj}^{H}+\bar{\rho}\sigma_{ID}^{2}\mathbf{I}_{R}\right)}.\label{eq:G_eye}
\end{equation}

\subsubsection{TWR-IA Algorithm}

Combining the above procedure, an iterative procedure is given in
Algorithm \ref{alg:2WRIA_algo}. 
\begin{algorithm}
\begin{algorithmic}[1]

\Require{ Initialize precoders $\mathbf{V}_{j},\forall j$, and
$\mathbf{G}=\mathbf{I}_{R}$.}

\For{ $t=1,2,\ldots,\text{max\_iter}$ }

\State{Get $\mathbf{G}=\alpha\mathbf{I}_{R}$ by \eqref{eq:G_eye}.}

\State{Obtain the combiner $\mathbf{U}_{k},\forall k$ via \eqref{eq:Uk}.}

\State{Compute the precoder $\mathbf{V}_{k},\forall k$ via \eqref{eq:Vk}.}

\EndFor

\end{algorithmic}

\caption{TWR-IA algorithm.\label{alg:2WRIA_algo}}
\end{algorithm}
After initializing precoders, the scalar $\alpha$, decoders and precoders
are iteratively updated until convergence. Since the scalar $\alpha$
is changed every iteration, the effective channel is also updated
per iteration. It can be seen via simulations that the number of iterations
for convergence depends on SNR. The higher the SNR, the larger the
number of iterations. Regarding the convergence, it can be seen that
the above problem can be shown jointly convex with respect to $\left(\mathbf{U}_{k},\mathbf{V}_{k},\forall k\right)$
as in \cite{9097459}. Both steps minimize the same MSE, and the MSE
is bounded to conclude the global convergence of the algorithm. 

\subsection{Proof of CD decomposition\label{sec:Proof-of-CD-1}}

\subsubsection{Lemma \ref{lem:MAT-CD-decomp-1}}

Consider two $M\times d$ orthonormal matrices $\mathbf{V},\hat{\mathbf{V}}$
such that $\mathbf{V}^{H}\mathbf{V}=\hat{\mathbf{V}}^{H}\hat{\mathbf{V}}=\mathbf{I}_{d}$.
Its left null space of size $M\times M-d$ can be represented as $\hat{\mathbf{V}}_{j}^{\text{null}}=\text{null}(\hat{\mathbf{V}}_{j})$.
Then, we can write 
\begin{align}
\mathbf{V} & =\hat{\mathbf{V}}\hat{\mathbf{V}}^{H}\mathbf{V}+\left(\mathbf{I}_{M}-\hat{\mathbf{V}}\hat{\mathbf{V}}^{H}\right)\mathbf{V}\nonumber \\
 & =\hat{\mathbf{V}}\underbrace{\hat{\mathbf{V}}^{H}\mathbf{V}}_{=\mathbf{X}\mathbf{Y}}+\hat{\mathbf{V}}^{\text{null}}\underbrace{\hat{\mathbf{V}}^{\text{null}H}\mathbf{V}}_{=\mathbf{S}\mathbf{Z}}
\end{align}
where the last equation is obtained by the QR-decomposition such that
$\mathbf{X}$ and $\mathbf{S}$ are $d\times d$ and $M-d\times d$
orthonormal matrices respectively. It verifies $d_{c}^{2}(\mathbf{V},\hat{\mathbf{V}})=d-\|\hat{\mathbf{V}}^{H}\mathbf{V}\|_{F}^{2}=d-\text{tr}(\mathbf{Y}^{H}\mathbf{Y})=\text{tr}(\mathbf{Z}^{H}\mathbf{Z})$.
Note that $\mathbf{X}\mathbf{Y}\in\mathbb{C}^{d\times d}$ is independent
of $\hat{\mathbf{V}}\in\mathbb{C}^{M\times d}$ since $\mathbf{XY}$
is a projection to a lower dimension space. Also, the factors $\mathbf{X}$
and $\mathbf{Y}$ are independent since $\mathbf{X}$ represents the
basis of $\hat{\mathbf{V}}^{H}\mathbf{V}$ and the basis is not unique.
Using similar facts, the matrices $\mathbf{S}$ and $\mathbf{Z}$
are also independent. For more details, visit \cite{4641957}. The
other two properties can be seen as follows. Consider the product
simplifications for $\hat{\mathbf{V}}^{\text{null}H}\hat{\mathbf{V}}^{\text{null}}=\mathbf{I}$
and $\hat{\mathbf{V}}^{\text{null}H}\hat{\mathbf{V}}=\mathbf{0}$
as 
\begin{align*}
\mathbf{I} & =\mathbf{V}^{H}\mathbf{V}\\
 & =\mathbf{Y}^{H}\mathbf{X}^{H}\hat{\mathbf{V}}^{H}\hat{\mathbf{V}}\mathbf{X}\mathbf{Y}+\mathbf{Z}^{H}\mathbf{S}^{H}\hat{\mathbf{V}}^{\text{null}H}\hat{\mathbf{V}}^{\text{null}}\mathbf{S}\mathbf{Z}\\
 & \qquad+2\Re\left(\mathbf{Z}^{H}\mathbf{S}^{H}\hat{\mathbf{V}}^{\text{null}H}\hat{\mathbf{V}}\mathbf{X}\mathbf{Y}\right)\\
 & =\mathbf{Y}^{H}\mathbf{X}^{H}\mathbf{X}\mathbf{Y}+\mathbf{Z}^{H}\mathbf{S}^{H}\mathbf{S}\mathbf{Z}=\mathbf{Y}^{H}\mathbf{Y}+\mathbf{Z}^{H}\mathbf{Z},
\end{align*}
and
\begin{align*}
d_{c}^{2}(\mathbf{V},\hat{\mathbf{V}}) & =d-\|\hat{\mathbf{V}}^{H}\mathbf{V}\|_{F}^{2}\\
 & =d-\|\hat{\mathbf{V}}^{H}\hat{\mathbf{V}}\mathbf{X}\mathbf{Y}+\hat{\mathbf{V}}^{H}\hat{\mathbf{V}}^{\text{null}}\mathbf{S}\mathbf{Z}\|_{F}^{2}\\
 & =d-\|\mathbf{X}\mathbf{Y}\|_{F}^{2}=d-\|\mathbf{Y}\|_{F}^{2}=\|\mathbf{Z}\|_{F}^{2}.
\end{align*}

\subsubsection{Corollary \ref{cor:CDD1}}

Let $\mathbf{V}_{j}$ and $\hat{\mathbf{V}}_{j}$ be two set of precoders
such that $d_{c}^{2}\left(\mathbf{V}_{j},\hat{\mathbf{V}}_{j}\right)=0,\forall j$,
i.e., from Lemma \ref{lem:MAT-CD-decomp-1}, $\mathbf{V}_{j}=\hat{\mathbf{V}}_{j}\mathbf{X}_{j}\mathbf{Y}_{j}$
with $\mathbf{X}_{j}\mathbf{X}_{j}^{H}=\mathbf{Y}_{j}\mathbf{Y}_{j}^{H}=\mathbf{I}_{d},\forall j$.
The sum rate and the harvested energy will be the same, since the
matrices with the zero CDs are related by a unitary matrix, which
cannot change the value of the products $\mathbf{V}_{j}\mathbf{V}_{j}^{H}=\hat{\mathbf{V}}_{j}\hat{\mathbf{V}}_{j}^{H},\forall j$,
the products $\bar{\mathbf{H}}_{kj}\bar{\mathbf{H}}_{kj}^{H},\forall j,k$
and the norm $\|\mathbf{H}_{rj}\mathbf{V}_{j}\|_{F}^{2},\forall j,k$.

\subsubsection{Corollary \ref{cor:Given-CD}}

From the CD decomposition, the desired displacement matrix can be
computed as $\mathbf{V}\bar{\mathbf{X}}\mathbf{Y}+\mathbf{V}^{\text{null}}\bar{\mathbf{S}}\mathbf{Z}$,
where $\bar{\mathbf{X}},\mathbf{Y},\bar{\mathbf{S}},\mathbf{Z}$ will
be computed to satisfy the constraint in Lemma \ref{lem:MAT-CD-decomp-1}.
The CD between this matrix and $\mathbf{V}$ can be written as 
\begin{align*}
z & =d_{c}^{2}\left(\mathbf{V}\bar{\mathbf{X}}\mathbf{Y}+\mathbf{V}^{\text{null}}\bar{\mathbf{S}}\mathbf{Z},\mathbf{V}\right)\\
 & \stackrel{(a)}{=}d_{c}^{2}\left(\mathbf{V}\bar{\mathbf{X}}\mathbf{U}_{Y}\Sigma_{Y}\mathbf{V}_{Y}^{H}+\mathbf{V}^{\text{null}}\bar{\mathbf{S}}\mathbf{U}_{Z}\Sigma_{Z}\mathbf{V}_{Y}^{H},\mathbf{V}\right)\\
 & \stackrel{(b)}{=}d_{c}^{2}\left(\mathbf{V}\mathbf{X}\Sigma_{Y}+\mathbf{V}^{\text{null}}\mathbf{S}\Sigma_{Z},\mathbf{V}\mathbf{V}_{Y}\right)\\
 & \stackrel{(c)}{=}d_{c}^{2}\left(\mathbf{V}\mathbf{X}\Sigma_{Y}+\mathbf{V}^{\text{null}}\mathbf{S}\Sigma_{Z},\mathbf{V}\right)=d_{c}^{2}\left(\mathbf{V}_{D},\mathbf{V}\right),
\end{align*}
where in $(a)$, the singular value decomposition (SVD) of $\mathbf{Z}=\mathbf{U}_{Z}\Sigma_{Z}\mathbf{V}_{Y}^{H}$
and $\mathbf{Y}=\mathbf{U}_{Y}\Sigma_{Y}\mathbf{V}_{Y}^{H}$ with
the same right singular vectors due to the constraint $\mathbf{Y}^{H}\mathbf{Y}=\mathbf{I}_{d}-\mathbf{Z}^{H}\mathbf{Z}$;
in $(b)$, $\mathbf{S}=\bar{\mathbf{S}}\mathbf{U}_{Z}$, $\mathbf{X}=\bar{\mathbf{X}}\mathbf{U}_{Y}$
are substituted, and the unitary matrix $\mathbf{V}_{Y}$ is multiplied
into both arguments, since the resulting CD is unchanged for unitary
multiplication, as in $(c)$. This shows that $\mathbf{Z}$ and $\mathbf{Y}$
can be relaxed to a diagonal matrix. 

\subsection{Proof of Lemma \ref{lem:Rate-loss-basic}: RLUB\label{subsec:Proof-of-LemmaRLUB}}
\begin{IEEEproof}
From the literature, we know that the rate loss is proportional to
the interference \cite{4641957,6117048}. Therefore, the rate loss
upper bound can be obtained as $\mathbb{E}\left\{ \Delta R_{k}\right\} \leq$
\begin{align*}
 & \frac{1}{2}\log_{2}\left|\mathbf{I}_{d}+\bar{\rho}\mathbb{E}\left\{ \left(\sum_{j\neq k,k'}\frac{P_{j}}{d}\underline{\mathbf{H}}_{kj}\hat{\mathbf{V}}_{j}\hat{\mathbf{V}}_{j}^{H}\underline{\mathbf{H}}_{kj}^{H}\right)\mathbf{N}_{k}^{-1}\right\} \right|,
\end{align*}
where in (a), $\underline{\mathbf{H}}_{kj}^{H}=\mathbf{U}_{k}^{H}\mathbf{H}_{kj}$,
and the inequality is obtained by Jensen's inequality. Now, using
the CD decomposition, one can write in terms of perfect IA precoder
as 
\begin{align*}
\bar{\mathbf{H}}_{kj}^{H}\hat{\mathbf{V}}_{j} & =\bar{\mathbf{H}}_{kj}^{H}\mathbf{V}_{j}\mathbf{X}_{j}\mathbf{Y}_{j}+\bar{\mathbf{H}}_{kj}^{H}\mathbf{S}_{j}\mathbf{Z}_{j}=\bar{\mathbf{H}}_{kj}^{H}\mathbf{S}_{j}\mathbf{Z}_{j},
\end{align*}
where $\underline{\mathbf{H}}_{kj}^{H}\mathbf{V}_{j}=0$ using IA.
In the above decomposition, $\mathbf{S}_{j}\in\mathcal{G}_{M\times d}$
and $\mathbf{Z}_{j}$ are independent of each other \cite[Lemma 1]{4641957}.
The above matrix $\bar{\mathbf{H}}_{kj}$ is not an orthonormal $M\times d$
matrix. To make it orthonormal, let $\tilde{\mathbf{H}}_{kj}=\underline{\mathbf{H}}_{kj}\mathbf{W}_{kj}\mathbf{\Lambda}_{kj}^{-1/2}$,
subject to $\tilde{\mathbf{H}}_{kj}^{H}\tilde{\mathbf{H}}_{kj}=\mathbf{I}_{d}$
and $\mathbf{W}_{kj}^{H}\mathbf{W}_{kj}=\mathbf{I}_{d}$, where $\underline{\mathbf{H}}_{kj}=\tilde{\mathbf{H}}_{kj}\mathbf{\Lambda}_{kj}^{1/2}\mathbf{W}_{kj}^{H}$
via singular-value decomposition and $\tilde{\mathbf{H}}_{kj}$, $\mathbf{W}_{kj}$,
and $\mathbf{\Lambda}_{kj}$ are independent of each other. Thus,
the following product is composed of independent terms which can be
simplified as 
\begin{align*}
 & \mathbb{E}\left\{ \underline{\mathbf{H}}_{kj}^{H}\mathbf{S}_{j}\mathbb{E}\left\{ \mathbf{Z}_{j}\mathbf{Z}_{j}^{H}\right\} \mathbf{S}_{j}^{H}\underline{\mathbf{H}}_{kj}\mathbf{N}_{k}^{-1}\right\} \\
\stackrel{(a)}{=} & \frac{z_{j}}{d}\mathbb{E}\left\{ \mathbf{W}_{kj}(\mathbf{\Lambda}_{kj}^{1/2})^{H}\tilde{\mathbf{H}}_{kj}^{H}\mathbf{S}_{j}\mathbf{S}_{j}^{H}\tilde{\mathbf{H}}_{kj}\mathbf{\Lambda}_{kj}^{1/2}\mathbf{W}_{kj}^{H}\mathbf{N}_{k}^{-1}\right\} \\
\stackrel{(b)}{=} & \mathbb{E}\left\{ \mathbf{W}_{kj}(\mathbf{\Lambda}_{kj}^{1/2})^{H}\mathbb{E}\left\{ \tilde{\mathbf{H}}_{kj}^{H}\mathbf{S}_{j}\mathbf{S}_{j}^{H}\tilde{\mathbf{H}}_{kj}\right\} \mathbf{\Lambda}_{kj}^{1/2}\mathbf{W}_{kj}^{H}\mathbf{N}_{k}^{-1}\right\} \\
\stackrel{(c)}{=} & \frac{z_{j}}{d}\frac{d}{M-d}\mathbb{E}\left\{ \mathbf{W}_{kj}\mathbf{\Lambda}_{kj}\mathbf{W}_{kj}^{H}\mathbf{N}_{k}^{-1}\right\} \\
= & \frac{z_{j}}{M-d}\mathbb{E}\left\{ \underline{\mathbf{H}}_{kj}^{H}\underline{\mathbf{H}}_{kj}\mathbf{N}_{k}^{-1}\right\} 
\end{align*}
where in (a), the decomposition of $\bar{\mathbf{H}}_{kj}=\tilde{\mathbf{H}}_{kj}\mathbf{\Lambda}_{kj}^{1/2}\mathbf{W}_{kj}^{H}$
has been substituted, and the expectation on $\mathbf{Z}_{j}$ is
carried out, which is approximated to be $\frac{1}{d}\mathbb{E}\left\{ \text{tr}(\mathbf{Z}_{j}\mathbf{Z}_{j}^{H})\right\} =\frac{z_{j}}{d}$
as \cite[App. B]{4641957}, with $z_{j}$ being the expected CD value;
in (b), independence of basis matrices $\tilde{\mathbf{H}}_{kj}$
and $\mathbf{S}_{j}$ is used; (c) is simplified from the fact that
the product of two orthonormal matrices $\tilde{\mathbf{H}}_{kj}^{H}\mathbf{S}_{j}$
is matrix-Beta distributed random variable $BETA(d,M-2d)$, which
has a mean of $\frac{d}{M-d}$. Next, we write for i.i.d. entries
in $\mathbf{H}_{kr}$ and $\mathbf{H}_{rj}$ as 
\begin{align*}
 & \mathbb{E}\left\{ \underline{\mathbf{H}}_{kj}^{H}\underline{\mathbf{H}}_{kj}\mathbf{N}_{k}^{-1}\right\} \\
= & \mathbb{E}\Big\{\mathbf{U}_{k}^{H}\mathbf{H}_{kr}\mathbf{G}\mathbf{H}_{rj}\mathbf{H}_{rj}^{H}\mathbf{G}^{H}\mathbf{H}_{kr}^{H}\mathbf{U}_{k}\\
 & \hfill\times\left[\bar{\rho}\sigma_{ID}^{2}\mathbf{U}_{k}^{H}\mathbf{H}_{kr}\mathbf{G}\mathbf{G}^{H}\mathbf{H}_{kr}^{H}\mathbf{U}_{k}+\sigma^{2}\mathbf{I}_{d}\right]^{-1}\Big\}\\
\stackrel{(a)}{=} & M\beta_{rj}\mathbb{E}\Big\{\mathbf{U}_{k}^{H}\mathbf{H}_{kr}\mathbf{G}\mathbf{G}^{H}\mathbf{H}_{kr}^{H}\mathbf{U}_{k}\Big\}\\
 & \hfill\times\left[\bar{\rho}\sigma_{ID}^{2}\mathbb{E}\mathbf{U}_{k}^{H}\mathbf{H}_{kr}\mathbf{G}\mathbf{G}^{H}\mathbf{H}_{kr}^{H}\mathbf{U}_{k}+\sigma^{2}\mathbf{I}_{d}\right]^{-1}\\
\stackrel{(b)}{\preceq} & M\beta_{rj}\mathbb{E}\Big\{\mathbf{U}_{k}^{H}\mathbf{H}_{kr}\mathbf{G}\mathbf{G}^{H}\mathbf{H}_{kr}^{H}\mathbf{U}_{k}\\
 & \hfill\times\left[\bar{\rho}\sigma_{ID}^{2}\mathbf{U}_{k}^{H}\mathbf{H}_{kr}\mathbf{G}\mathbf{G}^{H}\mathbf{H}_{kr}^{H}\mathbf{U}_{k}+\sigma^{2}\mathbf{I}_{d}\right]^{-1}\Big\}\\
\stackrel{(c)}{=} & \frac{M\beta_{rj}}{\bar{\rho}\sigma_{ID}^{2}+\frac{\sigma^{2}}{\beta_{kr}\mathbb{E}\|\mathbf{G}\|_{F}^{2}}}\mathbf{I}_{d}\\
\stackrel{(d)}{\approx} & \frac{M\beta_{rj}}{\bar{\rho}\sigma_{ID}^{2}+\left(\sum_{j=1}^{2K}\bar{\rho}P_{j}\beta_{rj}+\bar{\rho}\sigma_{ID}^{2}\right)\frac{\sigma^{2}}{\beta_{kr}P_{r}}}\mathbf{I}_{d}
\end{align*}
where in (a), $\mathbb{E}\mathbf{H}_{rj}\mathbf{H}_{rj}^{H}=M\beta_{rj}\mathbf{I}_{R}$;
in (b), the inequality $\mathbb{E}\left\{ \mathbf{C}_{x}\left(\mathbf{C}_{x}+\kappa\mathbf{I}\right)^{-1}\right\} \preceq\mathbb{E}\left\{ \mathbf{C}_{x}\right\} \left[\mathbb{E}\left\{ \mathbf{C}_{x}+\kappa\mathbf{I}\right\} \right]^{-1}$
is used (and verified via simulations); in (c), %
$\mathbb{E}\left[\mathbf{H}_{kr}\mathbf{G}\mathbf{G}^{H}\mathbf{H}_{kr}^{H}\right]=\beta_{kr}tr(\mathbf{G}\mathbf{G}^{H})\mathbf{I}_{M}$
and $\mathbf{U}_{k}^{H}\mathbf{U}_{k}=\mathbf{I}$ are used; in (d),
the fact that the norm of a vector is independent of its direction,
and the approximation $\frac{\mathbf{G}^{H}\mathbf{G}}{\|\mathbf{G}\|_{F}^{2}}\approx\frac{1}{R}\mathbf{I}_{R}$
are used to get $\mathbb{E}\|\mathbf{G}\|_{F}^{2}$ as%
{} $P_{r}=$
\begin{align*}
 & \mathbb{E}\|\mathbf{G}\|_{F}^{2}\mathbb{E}\text{tr}\left[\left(\sum_{j=1}^{2K}\mathbf{H}_{rj}\mathbf{V}_{j}\mathbf{V}_{j}^{H}\mathbf{H}_{rj}^{H}\bar{\rho}\frac{P_{j}}{d}+\bar{\rho}\sigma_{ID}^{2}\mathbf{I}_{R}\right)\frac{\mathbf{G}^{H}\mathbf{G}}{\|\mathbf{G}\|_{F}^{2}}\right]\\
 & \approx\mathbb{E}\|\mathbf{G}\|_{F}^{2}\mathbb{E}\text{tr}\left[\left(\sum_{j=1}^{2K}\mathbf{H}_{rj}\mathbf{V}_{j}\mathbf{V}_{j}^{H}\mathbf{H}_{rj}^{H}\bar{\rho}\frac{P_{j}}{d}+\bar{\rho}\sigma_{ID}^{2}\mathbf{I}_{R}\right)\frac{1}{R}\right]\\
 & =\mathbb{E}\|\mathbf{G}\|_{F}^{2}\left(\sum_{j=1}^{2K}d\bar{\rho}\frac{P_{j}}{d}\beta_{rj}+\bar{\rho}\sigma_{ID}^{2}\right)
\end{align*}
 Finally, the rate loss bound expression can be simplified as $\Delta R_{k}\leq$
\begin{align*}
 & \frac{1}{2}\log_{2}\left|\mathbf{I}_{d}+\frac{M_{d}\bar{\rho}\sum_{j\neq k,k'}P_{j}z_{j}\beta_{rj}\mathbf{I}_{d}}{\bar{\rho}\sigma_{ID}^{2}+\left(\sum_{j=1}^{2K}\bar{\rho}P_{j}\beta_{rj}+\bar{\rho}\sigma_{ID}^{2}\right)\frac{\sigma^{2}}{\beta_{kr}P_{r}}}\right|\\
 & =\frac{d}{2}\log_{2}\left(1+\frac{M_{d}\bar{\rho}\sum_{j\neq k,k'}P_{j}z_{j}\beta_{rj}}{\bar{\rho}\sigma_{ID}^{2}+\left(\sum_{j=1}^{2K}\bar{\rho}P_{j}\beta_{rj}+\bar{\rho}\sigma_{ID}^{2}\right)\frac{\sigma^{2}}{\beta_{kr}P_{r}}}\right),
\end{align*}
where $M_{d}=\frac{M}{d(M-d)}$. Substituting the value of $\bar{\rho}\sigma_{ID}^{2}=\bar{\rho}\sigma_{R}^{2}+\delta^{2}$
provides the required expression. %
\end{IEEEproof}

\subsection{Proof of Lemma \ref{lem:bounds}\label{subsec:Proof-of-bounds}}
\begin{IEEEproof}
The inequality in the upper bound comes from \eqref{eq:max_EH-Vj}
as $\frac{1}{d}\left\Vert \mathbf{H}_{rj}\mathbf{V}_{j}^{BAL}\right\Vert _{F}^{2}\leq\frac{1}{d}\sum_{i=1}^{d}\lambda_{ji}\leq\lambda_{j1},\forall j$,
where the inequality is due to the fact that the average of $d$-values
is less than the the maximum of them.

The inequality of the lower bound can be derived from the CD decomposition,
where the equality occurs when $z_{j}=0,\forall j$. For the proposed
balanced precoder with the optimum values of $\mathbf{X}_{j}^{*},\mathbf{Y}_{j}^{*},\mathbf{S}_{j}^{*}$
and $\mathbf{Z}_{j}^{*}$, we can write 
\begin{align*}
 & \left\Vert \mathbf{H}_{rj}\mathbf{V}_{j}^{BAL}\right\Vert _{F}^{2}\\
 & =\left\Vert \mathbf{H}_{rj}\mathbf{V}_{j}\mathbf{X}_{j}^{*}\mathbf{Y}_{j}^{*}+\mathbf{H}_{rj}\mathbf{V}_{j}^{\text{null}}\mathbf{S}_{j}^{*}\mathbf{Z}_{j}^{*}\right\Vert _{F}^{2}\\
 & \stackrel{(a)}{\geq}\left\Vert \mathbf{H}_{rj}\mathbf{V}_{j}\mathbf{X}_{j}^{*}\sqrt{1-\frac{z_{j}}{d}}+\mathbf{H}_{rj}\mathbf{V}_{j}^{\text{null}}\mathbf{S}_{j}^{*}\sqrt{\frac{z_{j}}{d}}\right\Vert _{F}^{2}\\
 & \stackrel{(b)}{\geq}\left\Vert \mathbf{H}_{rj}\mathbf{V}_{j}\mathbf{X}_{j}^{*}\right\Vert _{F}^{2}\left(1-\frac{z_{j}}{d}\right)+\left\Vert \mathbf{H}_{rj}\mathbf{V}_{j}^{\text{null}}\mathbf{S}_{j}^{*}\right\Vert _{F}^{2}\left(\frac{z_{j}}{d}\right)\\
 & \stackrel{(c)}{\geq}\left\Vert \mathbf{H}_{rj}\mathbf{V}_{j}\right\Vert _{F}^{2}\left(1-\frac{z_{j}}{d}\right)+\left\Vert \mathbf{H}_{rj}\mathbf{V}_{j}^{\text{n}}\right\Vert _{F}^{2}\left(\frac{z_{j}}{d}\right),
\end{align*}
where in $(a)$, the maximum value of the norm is upper bounded by
trivial selection $\mathbf{Z}_{j}=\mathbf{I}_{d}\sqrt{1-\frac{z_{j}}{d}}$;
in $(b)$, we employ the fact that the trace value in the expansion
of norm is non-negative for the proposed scheme, as mentioned in the
proposition \ref{prop:trace-val-noneg}; and in $(c)$, the specific
$d$-dimensional null space $\left(\mathbf{V}_{j}^{\text{null}}\mathbf{S}_{j}^{*}\right)$
can be replaced with any other $d$-dimensional null space $\mathbf{V}_{j}^{\text{n}}\in\mathcal{G}_{M,d}$
of $\mathbf{V}_{j}$. 
\end{IEEEproof}
\bibliographystyle{IEEEtran}
\bibliography{one}

\end{document}